\documentclass[journal]{IEEEtran}
\IEEEoverridecommandlockouts

\def\BibTeX{{\rm B\kern-.05em{\sc i\kern-.025em b}\kern-.08em
    T\kern-.1667em\lower.7ex\hbox{E}\kern-.125emX}}

\usepackage{cite}

\ifCLASSINFOpdf
   \usepackage[pdftex]{graphicx}
  \graphicspath{{./figures/}}
   \DeclareGraphicsExtensions{.pdf,.jpg,.JPG,.png,.eps}
\else
   \usepackage[dvips]{graphicx}
   \graphicspath{{./figures/}}
   \DeclareGraphicsExtensions{.eps}
\fi
\usepackage{amsmath}
\usepackage{multirow}
\usepackage{pbox}
\usepackage{array}

\usepackage[hyphens]{url}
\usepackage[hidelinks,bookmarks=false]{hyperref}
\hypersetup{breaklinks=true}
\usepackage{wasysym}
\usepackage{bm}
\usepackage{amssymb}
\usepackage{gensymb}
\usepackage{makecell}
\usepackage{siunitx}
\usepackage{fixltx2e}
\usepackage{xspace}
\usepackage{datetime}
\usepackage[dvipsnames]{xcolor}

\hypersetup{
    colorlinks=true,
    linkcolor=blue,
    citecolor=blue,
    urlcolor=blue
}

\usepackage{doi}
\usepackage{orcidlink}

\usepackage[framemethod=TikZ]{mdframed}

\mdfdefinestyle{MyFrame}{
    skipabove=4pt,
    skipbelow=4pt,
    innerrightmargin=4pt,
    innerleftmargin=4pt,
    leftmargin=2pt,
    rightmargin=2pt,
    backgroundcolor=black!15,
    hidealllines=true,
    roundcorner=2pt,
}

\begin{document}

\bstctlcite{IEEEexample:BSTcontrol}


\title{CXL-DMSim: A Full-System CXL\\Disaggregated Memory Simulator With Comprehensive Silicon Validation}
\author{Yanjing~Wang$^{\orcidlink{0009-0008-1466-308X}}$,
	Lizhou~Wu$^{\orcidlink{0000-0003-4439-7436}}$,
	Wentao~Hong,
	Yang~Ou,
	Zicong~Wang$^{\orcidlink{0000-0002-3106-5451}}$,
    Sunfeng~Gao$^{\orcidlink{0009-0008-6364-5638}}$,
    Jie~Zhang$^{\orcidlink{0000-0002-6299-4683}}$,
    Sheng~Ma$^{\orcidlink{0000-0003-1710-4060}}$,
    Dezun~Dong$^{\orcidlink{0000-0001-6243-8479}}$,
    Xingyun~Qi,
    Mingche~Lai$^{\orcidlink{0009-0001-5830-3974}}$,
    and~Nong~Xiao$^{\orcidlink{0000-0002-2166-977X}}$,~\IEEEmembership{Senior~Member,~IEEE}

	\thanks{Manuscript received 6 March 2025; revised 19 June 2025; accepted 2 September 2025. This work was supported in part by the National Natural Science Foundation of China under Grant 62304257 and 62332021; in part by the National Key Research and Development Program of China under Grant 2022YFB4500304. This article was recommended by Associate Editor Qinru Qiu. \textit{(Yanjing~Wang and Lizhou~Wu contributed equally to this work.) (Corresponding author: Mingche~Lai.)}}
	\thanks{Yanjing~Wang, Lizhou~Wu, Wentao~Hong, Yang~Ou, Zicong~Wang, Sunfeng~Gao, Sheng~Ma, Dezun~Dong, Xingyun~Qi, Mingche~Lai, and Nong~Xiao are with National University of Defense Technology, Changsha 410073, China (e-mail: \{wangyanjing, lizhou.wu, hongwt22, ouyang06, wangzicong, gaosunfeng20, masheng, dong, qi\_xingyun, mingchelai, nongxiao\}@nudt.edu.cn).} 
	\thanks{Jie~Zhang is with Peking University, Beijing 100871, China (e-mail: jiez@pku.edu.cn).}
	\thanks{Digital Object Identifier 10.1109/TCAD.2025.3607145} 
}

\markboth{IEEE TRANSACTIONS ON COMPUTER-AIDED DESIGN OF INTEGRATED CIRCUITS AND SYSTEMS, MANUSCRIPT FOR REVIEW, 2025}%
{Wang \MakeLowercase{\textit{et al.}}: A Full-System CXL Disaggregated Memory Simulator}
%


\maketitle

\begin{abstract}
Compute eXpress Link (CXL) has emerged as a key enabler of memory disaggregation for future heterogeneous computing systems to expand memory on-demand and improve resource utilization. However, CXL is still in its infancy stage and lacks commodity products on the market, thus necessitating a reliable system-level simulation tool for research and development. In this paper, we propose CXL-DMSim\footnote{Merged into SimCXL at \url{https://github.com/TianheMICALab/SimCXL}.}, an open-source full-system simulator to simulate CXL disaggregated memory systems with high fidelity at a gem5-comparable simulation speed. CXL-DMSim incorporates a flexible CXL memory expander model along with its associated device driver, and CXL protocol support with CXL.io and CXL.mem. It can operate in both app-managed mode and kernel-managed mode, with the latter using a dedicated NUMA-compatible mechanism. The simulator has been rigorously verified against a real hardware testbed with both FPGA- and ASIC-based CXL memory devices, which  demonstrates the qualification of CXL-DMSim in simulating the characteristics of various CXL memory devices at an average simulation error of 3.4\%. The experimental results using LMbench and STREAM benchmarks suggest that the CXL-FPGA memory exhibits a  $\sim$2.88$\times$  higher latency than  local DDR while the  CXL-ASIC latency is $\sim$2.18$\times$; CXL-FPGA achieves 45-69\% of local DDR memory bandwidth, whereas the number for CXL-ASIC is 82-83\%. The study also reveals that CXL memory can significantly enhance the performance of memory-intensive applications, improved by 23$\times$  at most with limited local memory for Viper key–value database and approximately 60\% in memory-bandwidth-sensitive scenarios such as MERCI. Moreover, the simulator's observability and expandability are showcased with detailed case-studies, highlighting its great potential for research on future CXL-interconnected hybrid memory pool.
\end{abstract}

\begin{IEEEkeywords}
Compute express link, Memory disaggregation, Full-system simulator, Hardware prototyping, Performance benchmarking.
\end{IEEEkeywords}

%
\IEEEpeerreviewmaketitle

\newpage
\section{Introduction}

\IEEEPARstart{W}{ith} the prevalence of massive data-driven applications such as AI/ML and big data analytics, the demand for larger memory is ever-increasing in today's heterogeneous parallel computing systems.
Over the past two decades, the CPU performance has been boosted dramatically thanks to Moore's law and multi/many-core scaling. However, the memory capacity and bandwidth per core have been decreasing, which apparently poses a bottleneck for system performance~\cite{Meta2021OCP}. 
In modern datacenters, the deployment unit is typically a monolithic server which contains closely-coupled computing and memory resources. This monolithic architecture for many years is always CPU-biased, leading to memory over-provision across the whole system. It has been observed that approximately 40\% of aggregated memory is unused most of the time in the hyperscale infrastructure at Azure and Google, primarily due to stranded memory (leftover memory when all cores of a server are rented out) and untouched memory (rented yet not actually used) \cite{li2023pond,GoogleV3}.
Considering the rising DRAM chip price in recent years, the under-utilization of memory resources becomes prohibitively expensive, which greatly boosts the TCO of today's datacenters \cite{li2023pond}.

With the advent of \textit{memory disaggregation} technologies, new solutions can be explored to tackle the memory wall and memory under-utilization challenges \cite{al2023memory}.
Memory disaggregation technologies decouple memory resources from CPUs, providing a feasible option for memory pooling. 
Conventionally, remote direct memory access (RDMA) technology is exploited to realize memory disaggregation \cite{Juncheng_NSDI2017,remote_region,aifm}. But RDMA is based on networking IO semantics which requires specialized NICs and software intervention, leading to a latency multiple orders of magnitude longer than that of local memory access.
In recent years, several low-latency and high-bandwidth cache-coherent interconnect protocols such as OpenCAPI, CCIX, CXL arise in industry, which have shown advantages over RDMA for memory disaggregation \cite{OpenCAPI,gen-z,ccix}. Among them, the CXL protocol \cite{CXL} is considered most promising and embraced by an increasingly number of semiconductor vendors worldwide.
With CXL, memory expansion becomes more flexible over the interconnect fabric while enabling coherent memory access via \emph{load/store} instructions. 
Furthermore, the CXL protocol is independent on the underlying memory technology, which can be DRAM, Flash, or even emerging non-volatile memories such as MRAM and RRAM.
This facilitates the construction of a unified heterogeneous memory pool for future energy- and cost-efficient computing systems.

Despite its attractive features disclosed in the protocol specifications,  CXL is still in its infancy stage and lacks commodity products on the market. As a result, the prior research work on CXL-based memory disaggregation is conducted based on four main methods: software-based emulation \cite{QEMU,mess}, software-based simulation \cite{gem5-CXL-github,DisaggSim,DRackSim,CXLMemSim_github}, hardware-based emulation \cite{NUMA_sim1,arif2022exploiting}, and hardware prototyping \cite{gouk2022direct,maruf2023tpp,sun2023demystifying}. However, the software-based emulation such as QEMU fails to model the physical characteristics and internal micro-architecture of real CXL memory devices. The simulation attempts of gem5-CXL and CXLMemSim are both immature and flawed; gem5-CXL fails to accurately model the CXL protocol behavior and to provide clear access interfaces, while CXLMemSim lacks full-system simulation capabilities and cycle-level fidelity resulting in limited functionality and poor usability.
The hardware-based emulation such as remote NUMA lacks CXL protocol support and there is a big difference in the memory access path and performance. 
As for the fourth method, there are currently no market-ready prototypes of CXL-based memory-disaggregated systems; CXL commodity products are also expensive to produce and purchase. 
Given the above limitations, there is a clear need for an accurate, cost-effective, and flexible tool for research on CXL-based disaggregated memory systems. 

In this paper, we present CXL-DMSim, a \textbf{full-system} \underline{CXL} 
\underline{D}isaggregated \underline{M}emory \underline{Sim}ulator for cycle-level simulation, architectural exploration, and evaluation of CXL-interconnected memory systems. CXL-DMSim is as easily configurable as the original gem5 simulator and fits to a variety of CXL devices. It has been rigorously verified and calibrated against a real-world CXL 1.1 testbed with both an in-house ASIC-based CXL memory expander and an FPGA-based CXL device prototype. 
To the best of our knowledge, CXL-DMSim is the first usable full-system disaggregated memory simulator.
The main contributions of this paper are listed below.

\begin{itemize}
\item A flexible device model of CXL memory expander (type-3 device) which currently supports DRAM, NVM and Flash as underlying storage media.
\item Currently CXL1.1+ supports for CXL.io and CXL.mem sub-protocols, which are used to enumerate, configure, and access our CXL memory device on CXL-DMSim.  
\item A driver for the device to operate in an application-managed mode and a NUMA-aware memory management mechanism to operate in a kernel-managed mode.
\item An extensive silicon measurement and evaluation that validates CXL-DMSim including performance tests, usability\&fidelity tests, real-world application tests, and observation\&expandability tests; this verifies the system's feasibility and offers guidelines for appropriate usage of CXL disaggregated memory.
\end{itemize}

\section{Background and Motivation}
\label{Sec:Background and Related works}
\subsection{CXL Protocol}
Compute eXpress Link (CXL) is a cache-coherent high-speed interconnect for CPUs, memory expanders, and accelerators \cite{CXL}. 
CXL features memory coherence between host memories and device-attached memories, enabling resource pooling and sharing for higher performance and reduced software complexity at lower cost.
It was first introduced by Intel in 2019 and since then has been widely embraced by the industry. As of today, CXL has undergone updates with three major versions: 1.0/1.1, 2.0, and 3.0/3.1/3.2.

\begin{figure}[t]
\centering
\includegraphics[width=0.45\textwidth]{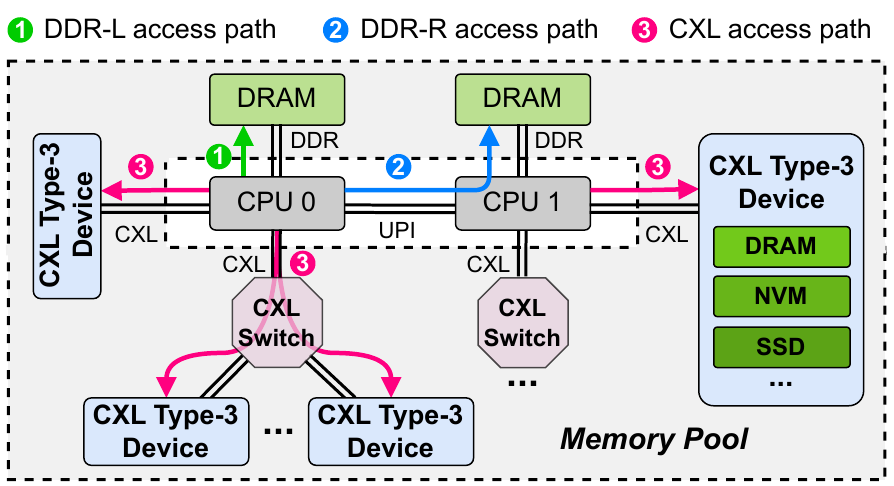}
\vspace{-5pt}
\caption{System memory expansion via CXL type-3 devices.} 
\label{fig:Expanding_memory_through_CXL}
\end{figure}

CXL is built on the PCIe physical layer and includes three sub-protocols: CXL.io, CXL.cache, and CXL.mem. CXL.io is functionally equivalent to PCIe in discovering and enumerating devices. CXL.cache enables I/O devices and accelerators to access and cache the host memory. CXL.mem allows the host to access memories that are attached to CXL devices via load/store instructions. The three sub-protocols identify three device types for distinct application scenarios. Type-1 devices employ CXL.io and CXL.cache sub-protocols. This type of devices are typically referred to as I/O devices such as SmartNIC which has no memory attached to it. Type-2 devices (accelerators), such as GPU with dedicated device memory and specialized compute units, utilize all three sub-protocols. Type-3 devices use the CXL.io and CXL.mem sub-protocols and are usually depicted as memory expanders, including both direct-attached and switch-attached ones, as shown in Fig.~\ref{fig:Expanding_memory_through_CXL}.
This paper focuses on type-3 devices for flexible memory expansion and pooling of future heterogeneous computing systems.

\vspace{-10pt}
\subsection{Related Work}
The prior work performs research on CXL-based memory disaggregation systems with the following four methods.

\subsubsection{\textbf{Software-Based Emulation}}
QEMU is a widely used and open-source emulator and virtualizer. As an emulator, it can perform full-system emulation which emulates all system components, including the processor and peripherals. 
In terms of CXL support, QEMU has begun to support CXL 2.0, with a few CXL components such as the CXL host bridge, CXL root ports, CXL switch, and CXL memory devices \cite{QEMU}. A recent work named Mess \cite{mess} presented a novel memory benchmark which measures bandwidth-latency curves. Based on the curves, a one-size-fit-all memory emulator was proposed to emulate the real-time latency of different types of main memories covering DDR4, DDR5, HBM, and CXL memories.

\subsubsection{\textbf{Software-Based Simulation}} 
gem5-CXL~\cite{gem5-CXL-github} is an open-source project which crudely adds a custom bus between the original gem5 MemBus and DDR controllers to introduce an extra delay. This extra delay actually models the deteriorated performance of CXL memory in comparison to DDR memory. DisaggSim \cite{DisaggSim} is similar to gem5-CXL but it does a step further by modeling a large-scale disaggregated memory system with Garnet model forming an interconnect fabric. DRackSim \cite{DRackSim} is an application-level simulation framework, which uses PinTool at the front-end to collect instruction traces from applications running on real machines. Its back-end performs cycle-level simulation based on these traces. 
CXLMemSim~\cite{CXLMemSim_github} is also a trace-driven simulator. Users can specify memory pool topologies and component latencies. It runs unmodified applications on real servers, divides execution into epochs, and recalculates execution time per epoch based on recorded memory accesses and modeled CXL latency.

\subsubsection{\textbf{Hardware-Based Emulation}} 
Due to the unavailability of commercial CXL hardware, many researchers have conducted emulations of CXL-expanded memory exploiting the NUMA mechanism. These studies \cite{NUMA_sim1,arif2022exploiting} leverage NUMA-enabled hardware to emulate CXL memory without directly handling the CXL protocol. Given the similar characteristics in bandwidth and latency between CXL memory and remote NUMA memory, this approach provides a reference for the study of computing systems with real CXL memory.

\subsubsection{\textbf{Hardware Prototyping}} 
There exist several ongoing projects to develop CXL hardware prototypes. For example, DirectCXL \cite{gouk2022direct} utilizes FPGAs to implement a CXL memory controller and a RISC-V processor that supports CXL. Transparent Page Placement (TPP)~\cite{maruf2023tpp} investigates page placement strategies for a tiered memory system using pre-production x86 CPUs with CXL 1.1 support and FPGA-based CXL memory expansion card. Sun et~al. \cite{sun2023demystifying} perform a comprehensive characterization of CXL memory on a real CXL hardware platform.

\subsection{Motivation}
Although QEMU serves as a generic system emulator, it is primarily used for functional-level emulation and verification. It does not model the physical characteristics and internal micro-architecture of  CXL devices. 
Thus, QEMU is unsuitable for architectural exploration and accurate performance evaluation of CXL-based systems. Similarly, the Mess emulator targets at reproducing the bandwidth-latency curves of the SystemC model of a Micron's CXL memory expander; real CXL memory devices may behave differently in a CXL-connected system. Moreover, Mess also lacks modeling of the CXL protocol, which limits its capability for architectural exploration especially when considering advanced CXL features such as  memory pooling and sharing.

Additionally, existing software-based simulation attempts are far away from readiness. gem5-CXL connects a simulated CXL device to the MemBus rather than mounting it on the IOBus to simulate peripherals. It does not model the behaviors of the CXL protocol (based on PCIe PHY) and lacks clear access interfaces with the modeled device. The project has not been updated for four years and is currently in a stagnant state.
DisaggSim and DRackSim perform coarse-grained modeling of CXL-enabled systems by abstracting key behaviors and modeling generalized components of disaggregated memory. However, these abstractions and simplifications introduce significant simulation errors.  Specifically, DisaggSim lacks calibration, whereas DRackSim is calibrated only against gem5 and public DirectCXL data, showing up to 12\% average error. Worse still, DRackSim’s custom host model runs 2$\times$ slower than gem5. CXLMemSim is not a full-system simulator in essence, as neither the CXL protocol and system architecture  nor the associated software stack are modeled. It lacks cycle accuracy and has limited functionalities, resulting in poor~usability.

As for the hardware-based emulation approach, the remote NUMA memory lacks realism and accuracy in mimicking CXL memory. Specifically, there is a significant difference in access latency and bandwidth  between these two types of memory, as emphasized in \cite{sun2023demystifying}; they also have different memory access paths (see Fig.~\ref{fig:Expanding_memory_through_CXL}).
In terms of the hardware prototype method, there are currently no market-ready hardware prototypes of CXL-based memory disaggregation systems. Existing hardware prototypes and commercial devices are expensive to produce, due to the long silicon development cycle.
Moreover, hardware prototypes also lack  flexibility in system configuration and observation.  

The limitations of current emulators, simulators and CXL hardware call for a  comprehensive and realistic simulation solution for agile design and evaluation of CXL-interconnected systems. This motivates us to propose  CXL-DMSim, which  is a  configurable, scalable, and cost-efficient simulator.

\begin{figure}[t]
\centering
\includegraphics[width=0.30\textwidth]{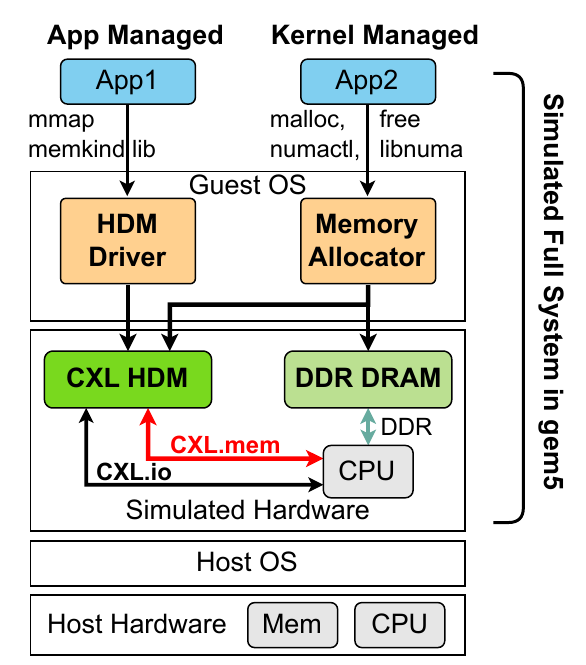}
\vspace{-5pt}
\caption{Architecture of CXL-DMSim simulator.} 
\label{fig:CXL_MemSim_overview}
\end{figure}

\section{CXL-DMSim Design and Implementation}
\label{sec:Design} 
\subsection{Simulator Architecture}
Fig.~\ref{fig:CXL_MemSim_overview} shows the overall architecture of CXL-DMSim, with three main components built on top of gem5. First, we added a CXL memory expander model, illustrated as CXL \emph{host-managed device memory} (HDM) in the figure, to the gem5 simulator. Second, we implemented the CXL sub-protocols CXL.io and CXL.mem, which are required for type-3 devices. The CXL.io sub-protocol is used for the CPU to enumerate and configure CXL devices; it is achieved by reusing the original PCI protocol in gem5. The CXL.mem sub-protocol allows the CPU to access the CXL HDM directly; it is achieved by integrating new CXL.mem packets as well as by extending internal components such as bridge in gem5. Third, we designed  a dedicated device driver and a NUMA-compatible memory management mechanism in the guest OS, to manage the allocation and deallocation of CXL HDM as well as to provide interfaces to upper-level user applications.

Depending on the CXL HDM management mechanism in the guest OS, CXL-DMSim provides users with two ways to use our CXL memory expander: \emph{application-managed} (AM) and \emph{kernel-managed} (KM). In the AM mode, applications can allocate and free CXL HDM using the \emph{memkind} library or the \emph{mmap} system call interface provided by our device driver. This usage mode offers users more flexibility and granularity in managing CXL HDM but also imposes the pressure of modifying legacy programs. In the KM mode, the CXL memory expander is exposed to the OS kernel as a CPU-less NUMA node, with the kernel managing the allocation and release of CXL HDM transparently to applications. Users can seamlessly utilize CXL HDM through existing tools such as \emph{numactl}. However, it is evident that in this mode the difference from the DDR memory is only reflected in the ``NUMA distance''. Users are not able to perceive and exploit the unique characteristics (e.g., persistence, large volume, and low power consumption) of different memory technologies behind the CXL interface. Note that the KM mode is consistent with the mainstream usage of CXL memory expanders on real-world hardware platforms.

\begin{figure}[t]
\centering
\includegraphics[width=0.50\textwidth]{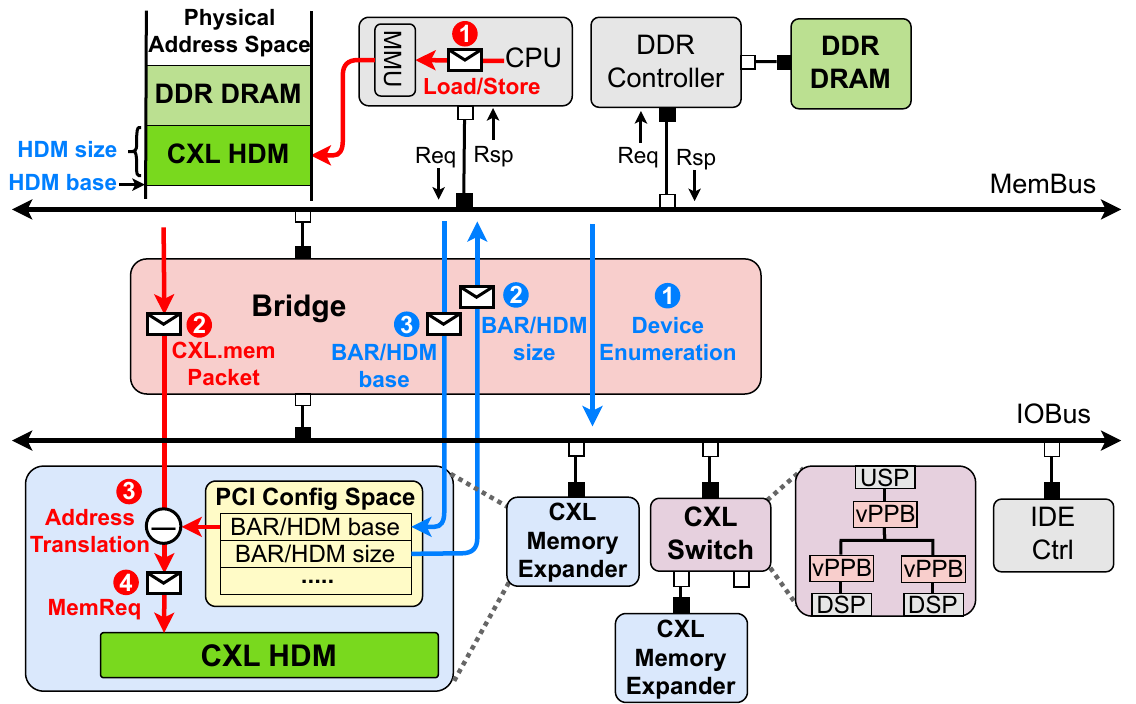}
\caption{System key component layout and CXL memory device access flow on the x86 platform of CXL-DMSim.} 
\label{fig:gem5_X86_platfrom}
\end{figure}

\vspace{-8pt}
\subsection{CXL Memory Expander Model}
gem5 is a modular, cycle-level computer system simulator. Its \emph{full-system} (FS) mode executes both user-level and kernel-level instructions and models a complete system including the OS and hardware devices, thereby simulating interactions between software and hardware more realistically \cite{gem5_paper,gem5_v20}. Fig.~\ref{fig:gem5_X86_platfrom} illustrates the structure of an x86 platform with the classic memory subsystem used in the FS simulation mode. Note that the operating flow marked with red and blue arrows will be explained in Sec.~\ref{Subsec:CXL_protocol_support}. The CPU, cache, and DDR controller are placed on the coherent MemBus, while devices such as PCI host and IDE controller are mounted on the non-coherent IOBus. These two buses are connected via a bridge. In our implementation, the CXL memory expander is accessed as a PCI device attached to the IOBus.

\begin{figure}[t]
\centering
\includegraphics[width=0.33\textwidth]{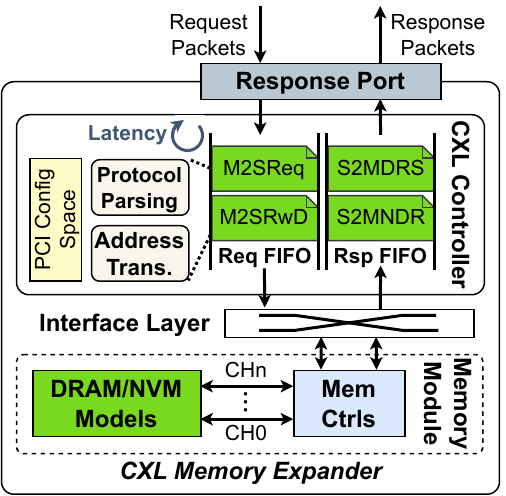}
\caption{CXL memory expander model design.} 
\label{fig:device_model}
\end{figure}

Fig.~\ref{fig:device_model} shows our CXL memory expander model in CXL-DMSim. The model is organized into four components from top to bottom: 1) a response port for communication, 2) a CXL controller for parsing read/write packets, 3) a flexible interface layer for request forwarding, and 4) a memory module including memory controllers and memory medium for data storage. Since native gem5 models external devices in a coarse-grained manner using the \textit{atomic mode} (transactional behavior is represented as simple accumulated latency), this approach loses critical timing information essential for memory modeling. To achieve accurate memory access behavior, \textbf{we introduce an event-driven \textit{timing mode} into the CXL memory expander}. The response port is connected to the IOBus, enabling it to receive request packets directed to the memory expander and send response packets to the requesters.

The CXL controller in the device model incorporates Req FIFO and Rsp FIFO buffers to manage incoming request packets from the port and response packets from the backend memory medium, respectively (note that the definitions of M2SReq, M2RRwD, S2MDRS, and S2MNDR packets shown in Fig.~\ref{fig:device_model} will be explained in Sec.~\ref{Subsec:CXL_protocol_support}). The CXL controller performs two critical operations on buffered packets. First, it parses CXL transaction packets, which include read and write configuration packets for CXL.io and memory transaction packets for CXL.mem. Second, it translates the address in the incoming packets to the actual address for accessing the backend memory controller. For the CXL.io sub-protocol, we have implemented the PCI configuration space of the expander. This includes basic device information and \textit{base address registers} (BARs), thereby enabling the device to respond to read and write transactions from the host to the configuration space. For the CXL.mem sub-protocol, the CXL controller first identifies the command fields in the packets according to the protocol specifications. Then, it calculates the actual address based on the destination address of the packet and the base address stored in the BARs. The derived read/write commands and their associated addresses are encapsulated into new requests, which are forwarded to the memory controller through the interface layer to execute load/store operations. This flexible interface layer decouples the frontend CXL controller from the backend memory module, enabling convenient customization of CXL memory characteristics such as latency, bandwidth, and volatility through memory medium replacement and adjustment of memory channel count.

The memory module is designed to be technology-agnostic, which means it can be the conventional charge-based memories such as DRAM and Flash, or emerging resistance-based non-volatile memories such as RRAM and MRAM. The memory capacity of the device is declared to the host via the BARs during system startup. When the memory controller's transaction queue receives read and write requests from the CXL controller, it schedules the requests to access data at the corresponding addresses and generates responses if needed. Thanks to our well-designed modular structure, existing memory models from gem5, including native DRAM models, DRAMSim models, and NVM models, can be directly integrated into our expander's backend. Additionally, we provide a coarse-grained DRAM model that offers faster simulation speed at the cost of reduced accuracy.

To capture the timing behavior of CXL memory expanders, our device model provides three key parameters: \textit{dev\_proto\_lat} regulates the processing latency of the CXL.mem sub-protocol, while \textit{dev\_req\_fifo\_depth} and \textit{dev\_rsp\_fifo\_depth} determine the CXL controller's capacity to handle concurrent packets via FIFO depth configuration. In the timing mode, the overall memory access latency and expander bandwidth are functions of packet queuing status, memory controller scheduling policies, and the temporal adjacency of memory requests.

\vspace{-5pt}
\subsection{CXL Protocol Support}
\label{Subsec:CXL_protocol_support}
The CXL memory expander uses CXL.io and CXL.mem sub-protocols. 
Since CXL.io is functionally similar to the traditional PCI protocol at the transaction layer, we enhanced the existing PCI protocol to implement device enumeration and configuration. This process is handled with three main steps during system startup, as shown with the blue arrows in Fig.~\ref{fig:gem5_X86_platfrom}. Step \raisebox{.5pt}{\textcircled{\raisebox{-.9pt} {1}}}: When enumerating a CXL memory expander, its driver in OS first queries  the internal memory size declared by the BARs using CXL.io configuration read transactions. Step \raisebox{.5pt}{\textcircled{\raisebox{-.9pt} {2}}}: Based on the retrieved size, the HDM is then mapped into the host's physical address space. Step \raisebox{.5pt}{\textcircled{\raisebox{-.9pt} {3}}}: The mapped base address is written into the expander configuration space via CXL.io configuration write transactions. By means of the above process, the CXL memory expander can be discovered and subsequently accessed by the host in the system. 

The host accesses HDM via the CXL.mem with four main steps, as illustrated with the red arrows in Fig.~\ref{fig:gem5_X86_platfrom}. Step \raisebox{.5pt}{\textcircled{\raisebox{-.9pt} {1}}}: Memory access requests are first initiated towards HDM through load/store instructions. Step \raisebox{.5pt}{\textcircled{\raisebox{-.9pt} {2}}}: These requests are then routed to downstream devices via the bridge, where gem5-internal packets are converted into CXL.mem packets in order to communicate with the CXL memory expander. Step \raisebox{.5pt}{\textcircled{\raisebox{-.9pt} {3}}}: When the CXL.mem packets arrive at the expander, the CXL controller converts the request address to the expander's internal memory address based on the HDM base address. Step \raisebox{.5pt}{\textcircled{\raisebox{-.9pt} {4}}}: The controller forwards the converted request to the backend memory controller for actual memory access.

\begin{figure}[t]
\centering
\includegraphics[width=0.35\textwidth]{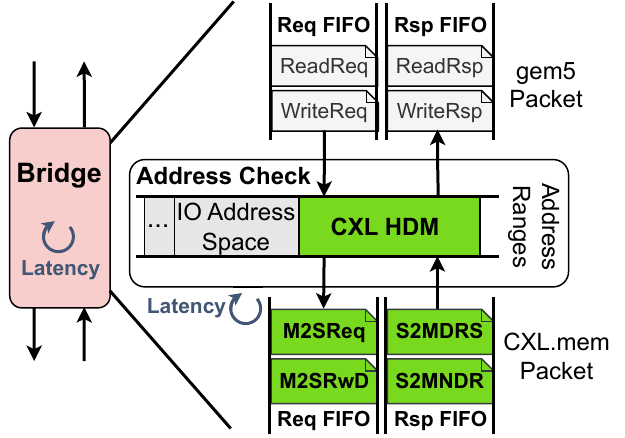}
\caption{Packet interception and transformation.} 
\label{fig:bridge_protocol_process}
\end{figure}

The CXL.mem sub-protocol defines two types of communication endpoints. One is called master, typically a local agent in the host processor. The other is subordinate such as a memory controller in the CXL memory expander. This protocol specifies two types of packets for communication between the master and subordinate agents: those from the master to the subordinate are called \textit{master-to-subordinate} (M2S) packets, and those in the opposite direction are called \textit{subordinate-to-master} (S2M) packets~\cite{CXL}. CXL-DMSim extends the packet-based point-to-point communication mechanism in gem5 to simulate the CXL.mem sub-protocol. First, when the CPU issues a ReadReq/WriteReq gem5 packet, its command field is updated to include a read/write transaction command used by the CXL.mem. Subsequently, the bridge connecting the memory bus and I/O bus intercepts the packet if destined to the CXL memory expander. Fig.~\ref{fig:bridge_protocol_process} shows that the bridge contains two pairs of FIFO queues, one pair for buffering upstream gem5 packets and the other for buffering downstream CXL.mem packets. The bridge checks the address and captures the gem5 packet in the upstream Req FIFO, then transforms it into a M2SReq/M2SRwD packet. Thereafter, the CXL.mem packet is buffered in the downstream Req FIFO and then sent out to the CXL memory expander. Similarly, when a response packet (S2MDRS/S2MNDR) from the CXL memory expander arrives at the bridge, it is buffered in the Rsp FIFO and then converted into a ReadRsp/WriteRsp packet. The gem5 packet is buffered in the upstream Rsp FIFO before being transmitted to the memory bus. 

We also implement a CXL switch model to support the single-level switching feature introduced in CXL2.0 protocol. With CXL switches, a larger disaggregated memory system can be built as illustrated Fig.~\ref{fig:gem5_X86_platfrom}.
In single-host configurations, a CXL switch contains a single upstream port (USP) and several downstream ports (DSPs). Each port corresponds to a virtual PCI-to-PCI bridge (vPPB), and these vPPBs are interconnected through a virtual PCI bus for packet forwarding. We model the CXL switch by extending the Bridge component. When the USP receives an upstream packet, its vPPB inspects the packet’s address and forwards the packet to the corresponding DSP. Similarly, a packet routed from an endpoint device to the host goes through the switch from the device's associated DSP to the USP in a reverse direction.

Note that the bridge module includes four configuration parameters:
The \textit{bridge\_lat} represents the inherent latency through the bridge module, which is already present in the original gem5 code. The \textit{host\_proto\_lat} represents the processing latency for CXL.mem sub-protocol packets. The \textit{link\_req\_fifo\_depth} and \textit{link\_rsp\_fifo\_depth} determine the depths of the FIFO queues used for buffering CXL.mem request and response packets, respectively. These queue depths impact the throughput of the CXL link. Similarly, the CXL switch includes two configurable parameters: \textit{switch\_lat} represents the packet forwarding latency through the switch, and \textit{switch\_buffer\_size} represents the size of the switch buffer.

\begin{figure}[t]
\centering
\includegraphics[width=0.48\textwidth]{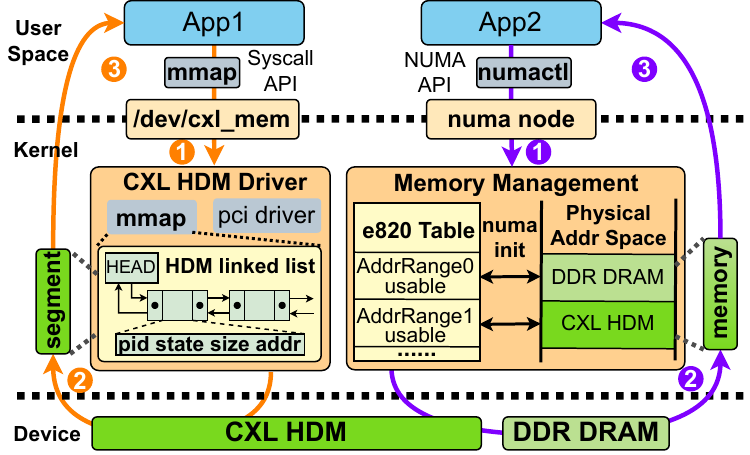}
\caption{Two ways of managing CXL HDM in the OS.} 
\label{fig:OS_design}
\end{figure} 

\vspace{-5pt}
\subsection{CXL HDM Management in OS}
Fig.~\ref{fig:OS_design} shows the software stacks of operating the CXL memory expander in both the AM and KM modes. In the AM mode, the CXL HDM driver serves not only for device enumeration and configuration but also furnishes upper-layer applications with a suite of system call interfaces to access and manage HDM. As the memory expander is mounted on the PCI bus as an external device to the host, its driver adheres to the PCI driver programming paradigm. During the enumeration phase of the system, the ``pci driver'' component of the driver identifies the CXL memory expander and creates a character device file named \emph{/dev/cxl\_mem}. The ``file operations'' component of the driver implements system calls (e.g., mmap) for accessing HDM and initializes a doubly linked list for managing HDM allocation. Each node in the linked list records the process \textit{PID} that owns the memory block, the current \textit{state} of the memory block (FREE or BUSY), the \textit{size} of the memory block, and the offset \textit{address} of the memory block.
An application accesses HDM via the AM mode in three main steps, as shown with the orange arrows in Fig.~\ref{fig:OS_design}. Step \raisebox{.5pt}{\textcircled{\raisebox{-.9pt} {1}}}: The application opens the device file \emph{/dev/cxl\_mem} using the driver's \emph{open} system call to acquire the device file descriptor, then calls the \emph{mmap} to request a fixed-size HDM region. Step \raisebox{.5pt}{\textcircled{\raisebox{-.9pt} {2}}}: The driver uses a mutex to ensure exclusive HDM access. Once the process gets the mutex, the allocator applies a best-fit algorithm on the HDM linked list to reserve a contiguous physical segment. If successful, it obtains the segment’s physical offset, maps the virtual address space to it, and updates the linked list. Step \raisebox{.5pt}{\textcircled{\raisebox{-.9pt} {3}}}: The application can use the pointer returned by the \emph{mmap} to read from or write to the HDM directly. Furthermore, the pointer can also be passed to the \emph{memkind} library for the purpose of unified memory management of DDR DRAM and CXL HDM.

Although the AM mode provides a straightforward method to access the CXL HDM, it requires modifications to applications' source code and does not utilize the kernel's existing tiered memory management infrastructure. As a result, the prevalent approach in real-world hardware platforms treats the CXL memory device as a CPU-less NUMA node. This approach remains transparent to applications and enables NUMA-aware memory management. Therefore, CXL-DMSim also provides a NUMA API for accessing HDM, which corresponds to the KM mode. To achieve this, we added an e820 table entry mapping to the CXL HDM, enabling the kernel to recognize the CXL HDM as part of the available system memory during boot time. Note that this approach mirrors the configuration process used on real hardware platforms for recognizing and configuring CXL devices. Additionally, we adjusted the kernel \emph{numa\_init} routine to initialize both DDR DRAM and CXL HDM as two separate NUMA nodes. An application accesses HDM via KM mode in three main steps, as shown with the purple arrows in Fig.~\ref{fig:OS_design}. Step \raisebox{.5pt}{\textcircled{\raisebox{-.9pt} {1}}}: The application first uses the \emph{numactl} tool to bind memory allocations to the HDM node or opt for other NUMA strategies, such as memory interleaving. Step \raisebox{.5pt}{\textcircled{\raisebox{-.9pt} {2}}}: The kernel's memory management subsystem then handles memory allocation transparently, with the allocated physical memory varying based on the specified NUMA strategy. Step \raisebox{.5pt}{\textcircled{\raisebox{-.9pt} {3}}}: The application can access HDM through standard memory allocation interfaces such as \emph{malloc}; the kernel is responsible for the allocation of memory pages across DDR DRAM and CXL HDM.

\vspace{-5pt}
\section{Experiments and Evaluation}
\label{sec:Experiments and Evaluation}

\begin{figure}[t]
\centering
\rotatebox{90}{
\includegraphics[width=0.29\textwidth]{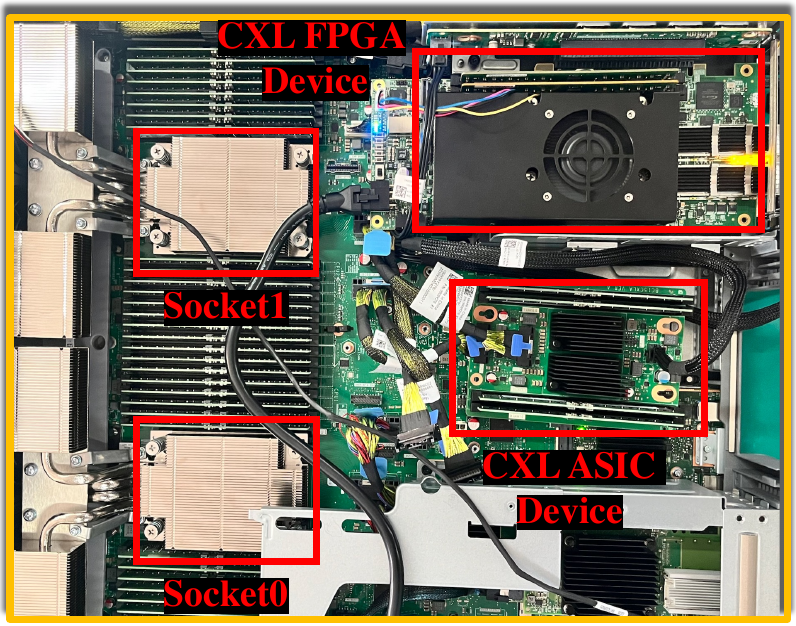}}
\vspace{-5pt}
\caption{Top view of our CXL hardware testbed.} 
\label{fig:hardware_prototype}
\end{figure}

\subsection{Experimental Setup}
We have conducted comprehensive experiments on both real hardware testbed and our proposed CXL-DMSim. The detailed configurations of these two test platforms are as follows. 

\subsubsection{\textbf{Hardware Testbed}}
\begin{figure}[t]
\centering
\includegraphics[width=0.48\textwidth]{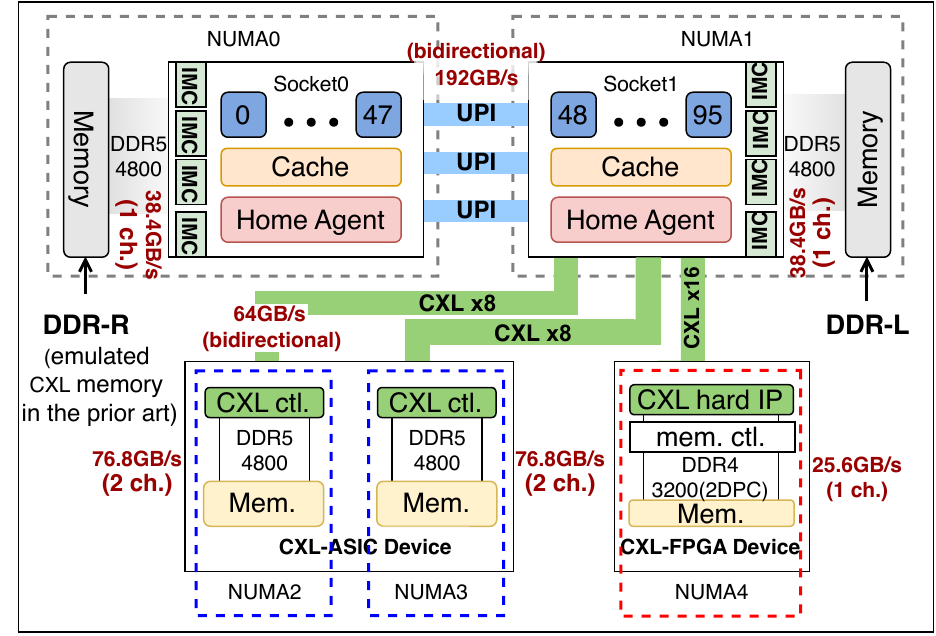}
\vspace{-5pt}
\caption{Architecture diagram of our CXL hardware testbed.} 
\label{fig:topology}
\vspace{-10pt}
\end{figure}

The testbed is a high-performance dual-socket x86 server shown in Fig.~\ref{fig:hardware_prototype}; its internal structure is illustrated in Fig.~\ref{fig:topology}. The server has dual-socket Intel Xeon Platinum 8468V processors, each of which has 4 integrated memory controllers with 8 memory channels. For the purpose of  fair comparison with a CXL link, we only enabled a single memory channel (i.e., a 32GB DDR5 4800MT/s DIMM) for each NUMA node in our characterization experiments. The detailed host configurations are listed in Table~\ref{tab:host_config}.

\begin{table}[t]
\centering
\renewcommand{\arraystretch}{1.0} 
\caption{Host configurations for our hardware testbed and CXL-DMSim simulator.}
\label{tab:host_config}
\resizebox{0.49\textwidth}{!}{
\begin{tabular}{|l|l|l|}
\hline
\textbf{Config. Parameter}  & \textbf{CXL Testbed Host} & \textbf{CXL-DMSim Host}     \\ 
\hline
{Linux kernel version} & {v6.5.0}  &{Modified v5.4.49} \\ \hline
{CPU type} &{Xeon® Platinum 8468V } & {X86O3CPU}  \\ \hline
{CPU cores}       &{48}       &  {48} \\ \hline
{Local DRAM type} &{DDR5 4800}& { DDR5 4400}    \\ \hline
{\#Memory channels}   &{1}    & {1} \\ \hline
{Local DRAM size}  &{32GB}   & {32GB}   \\ \hline
{L1 dcache size}&{48KB}  & {48KB} \\ \hline
{L1 icache size} &{32KB}& {32KB}  \\ \hline
{L2 cache size} &{2MB}& {2MB}  \\ \hline
{LLC size}& {97.5MB}  & {96MB} \\ \hline
\end{tabular}
}
\vspace{-10pt}
\end{table}

\begin{table}[t!]
\centering
\renewcommand{\arraystretch}{1.0} 
\caption{Real CXL memory expander configurations for\\CXL-FPGA and CXL-ASIC devices.}
\label{tab:cxl_config}
\resizebox{0.49\textwidth}{!}{
\begin{tabular}{|l|l|l|}
\hline
\textbf{Config. Parameter}  & \textbf{CXL-FPGA }  & \textbf{CXL-ASIC }   \\ 
\hline
{CXL memory size}   &  {16GB}   &{64GB}\\ \hline
{\#Backend memory channels}& {1}  & {2} \\ \hline
{Type of backend DRAM} & {DDR4 3200}& {DDR5 4800} \\ \hline
{Latency}& {375ns} & {284ns}   \\ \hline
\end{tabular}
}
\end{table}

On our hardware testbed, we measured two types of CXL memory devices:  a real CXL FPGA memory device (denoted as CXL-FPGA) and a real CXL ASIC memory device (denoted as CXL-ASIC); the details are listed in Table~\ref{tab:cxl_config}. Both of these two devices are connected to Socket1, as illustrated in Fig.~\ref{fig:topology}. 
The CXL-FPGA device is an Intel Agilex I-Series FPGA Development Kit with an integrated hard CXL IP~\cite{IntelFPGA}. It is equipped with 16GB 3200MT/s DDR4 memory, connected to the CPU via a CXL 1.1 interface. The CXL-FPGA device is recognized as a separate CPU-less NUMA node (denoted as NUMA4); it can be used and managed using existing NUMA software infrastructure. 
The in-house CXL-ASIC device has two CXL memory controllers, each of which is composed of a \textit{CXL protocol controller plus a DDR5 controller} driving two DIMMs. As each DIMM is populated with a 32GB DDR5 4800MT/s  memory module, the CXL-ASIC device owns a total memory capacity of 128GB. 
It appears in the OS as two separate NUMA nodes without CPUs (denoted as NUMA2 and NUMA3 respectively).

We also measured the DDR5 memory on the remote NUMA node (denoted as DDR-R on NUMA0) as an \textit{emulated CXL memory} device, as done in many of the prior works \cite{NUMA_sim1,arif2022exploiting}. 
For better comparing the performance of the above-mentioned three types of memory, we also measured the local DDR5 memory (denoted as DDR-L) on NUMA1 as a baseline.

\subsubsection{\textbf{CXL-DMSim Simulator}}
CXL-DMSim is implemented based on gem5 v23.1 in the full-system mode. The CPU type is X86O3CPU with 32GB 4400MT/s single DDR5 memory (denoted as CXL-DMSim\textsubscript{L}). 
The CXL memory device driver was developed based on the Linux v5.4.49 kernel. CXL-DMSim simulates a CXL FPGA memory device (denoted as CXL-DMSim\textsubscript{F}) with 16GB, a CXL ASIC memory device (denoted as CXL-DMSim\textsubscript{A}) with 64GB, and a CXL ASIC memory device interconnected via a switch (denoted as CXL-DMSim\textsubscript{S}), with configurations listed in Table~\ref{tab:cxl_dmsim_config}. To simulate different CXL memory devices from different vendors, we introduce nine configurable parameters and their typical ranges in our simulator, according to the latency breakdown of memory access paths provided in the CXL spec. \cite{CXL} and relevant literature \cite{DasSharma2023AnIT,li2023pond,Low_overhead,gouk2022direct,cxl_switch}. For our CXL FPGA and ASIC devices, the values of these parameters in Table~\ref{tab:cxl_dmsim_config} have been meticulously calibrated and fine-tuned using end-to-end latency measurements on our hardware testbed. For the switch-attached CXL ASIC device, we calibrated the switch's parameters to match the performance measurements reported by XConn \cite{cxl_switch}.
For those who are interested in using our simulator, different values can  be configured to match the performance of their own CXL memory devices.

\begin{table}[t]
\centering
\renewcommand{\arraystretch}{1} 
\caption{Modeled CXL memory expander configurations on CXL-DMSim for FPGA type (CXL-DMSim\textsubscript{F}), ASIC type (CXL-DMSim\textsubscript{A}) and switch-attached ASIC type (CXL-DMSim\textsubscript{S}).}
\label{tab:cxl_dmsim_config}
\resizebox{0.49\textwidth}{!}{
\begin{tabular}{|l|l|l|l|}
\hline
\textbf{Config. Parameter}  & \textbf{CXL-DMSim\textsubscript{F}}  & \textbf{CXL-DMSim\textsubscript{A}} & \textbf{CXL-DMSim\textsubscript{S}} \\ \hline
{CXL memory size}&{16GB}&{64GB}&{64GB}\\ \hline
{Type of backend DRAM} & {DDR4 3200}& {DDR5 4400}& {DDR5 4400} \\ \hline
{bridge\_lat}& {50ns}  & {50ns}& {50ns} \\ \hline
{host\_proto\_lat} & {12ns}& {12ns}& {12ns} \\ \hline
{dev\_proto\_lat}& {60ns} & {15ns}& {15ns}   \\ \hline
{link\_req\_fifo\_depth} & {128}  & {128}& {128}\\ \hline
{link\_rsp\_fifo\_depth}  & {128} & {128}& {128} \\ \hline
{dev\_rsp\_fifo\_depth}  & {36} & {48}& {48} \\ \hline
{dev\_rsp\_fifo\_depth}  & {36} & {48}& {48} \\ \hline
{switch\_lat}  & {N/A} & {N/A}& {100ns} \\ \hline
{switch\_buffer\_size}  & {N/A} & {N/A}& {64} \\ \hline
\end{tabular}
}
\end{table}

\vspace{-5pt}
\subsection{Usability and Fidelity of CXL-DMSim}
There exist many popular benchmarks to characterize memory performance in industry, such as Intel's MLC \cite{MLC}, LMbench \cite{lmbench}, STREAM \cite{stream}. Considering ease of use and compatibility on both hardware and simulator platforms, we selected LMbench, STREAM, and Redis-YCSB \cite{Redis, ycsb_cooper2010benchmarking} to evaluate the latency, bandwidth, and real-world application performance of DDR-L, DDR-R, CXL-FPGA, CXL-ASIC, and CXL-DMSim\textsubscript{F}, CXL-DMSim\textsubscript{A}, and CXL-DMSim\textsubscript{S}.
\label{subsec:circuit_sim}

\subsubsection{\textbf{LMbench Test for Memory Latency}}
To measure the access latency of CXL HDM, we utilized a memory latency benchmark \emph{lat\_mem\_rd} in LMbench \cite{lmbench}. It can be used to measure the random read latency of different memory layers, covering L1, L2 and L3 caches, as well as DDR-L, DDR-R and CXL memories. The random read pattern typically reflects the system's real latency \cite{mess}. Hardware prefetch mechanisms are all disabled in BIOS settings to provide more accurate measurements of memory latencies. Note that we only present the read latency as the CXL memory is symmetric in read and write performance. The benchmark involves two key parameters: array size and stride. To ensure that the entire memory hierarchy is accurately measured, avoiding the disturbance of caching on the results meanwhile properly representing the data access patterns in real-world applications, the array size should be set to at least four times as large as LLC and no more than 80\% of physical memory size \cite{lmbench}. Given the above considerations, we set the array size to \SI{2048}{MB} and the stride to 64 bytes in our experiments.

\begin{figure}[t]
\centering
\includegraphics[width=0.5\textwidth]{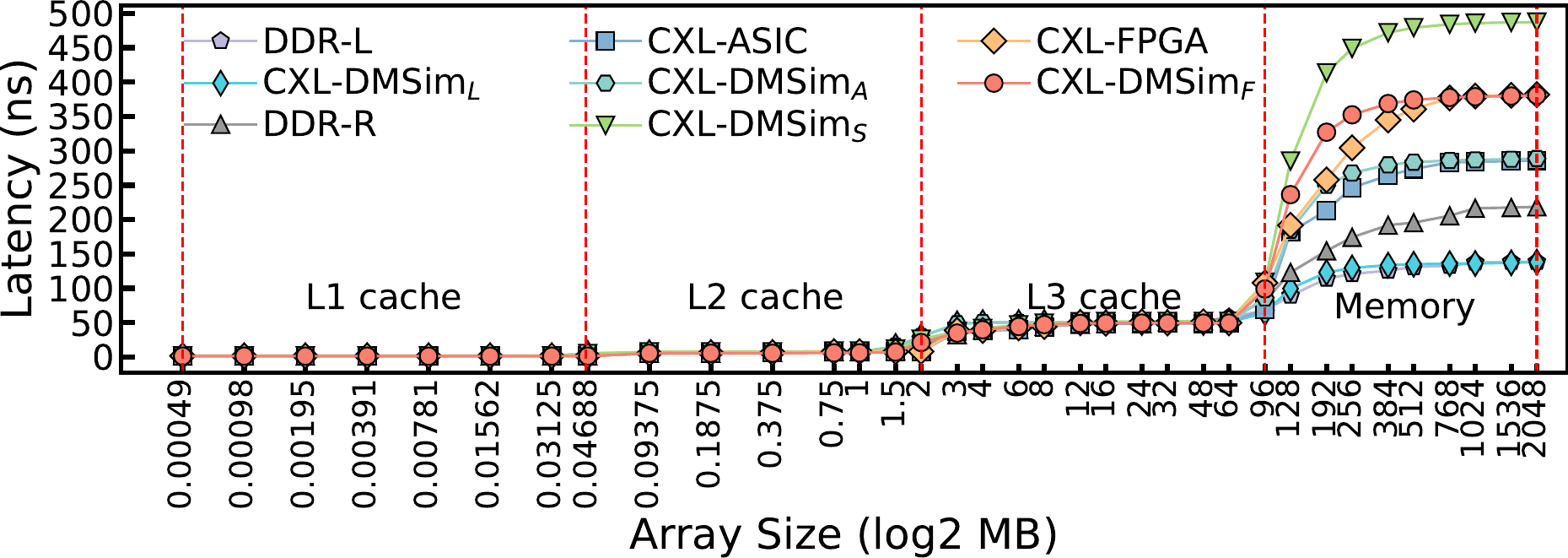}
\caption{Measured random memory access latency of eight memory devices: DDR-L, DDR-R, CXL-ASIC, CXL-FPGA, CXL-DMSim\textsubscript{L}, CXL-DMSim\textsubscript{A}, CXL-DMSim\textsubscript{F}, and CXL-DMSim\textsubscript{S}.} 
\label{fig:lmbench_latency}
\end{figure}

Fig.~\ref{fig:lmbench_latency} presents the measurement results using LMbench. The curves contain a series of plateaus, each of which represents a level in the memory hierarchy. The right-most plateau represents the random read latency of DDR-L (\SI{130}{\nano\second}), DDR-R (\SI{200}{\nano\second}), CXL-ASIC (\SI{284}{\nano\second}), CXL-FPGA (\SI{375}{\nano\second}), CXL-DMSim\textsubscript{L} (\SI{130}{\nano\second}), CXL-DMSim\textsubscript{A} (\SI{284}{\nano\second}),  CXL-DMSim\textsubscript{F} (\SI{375}{\nano\second}), and CXL-DMSim\textsubscript{S} (\SI{487}{\nano\second}).
The latency of CXL-ASIC is approximately 2.18 times higher than that of DDR-L, while the latency of CXL-FPGA is about 2.88 times. The two types of CXL memory devices exhibit much higher latencies because their access path is much longer than that of DDR-L. The latency of CXL-FPGA is \SI{91}{\nano\second} higher than CXL-ASIC, because FPGA-based CXL memory fails in fully utilizing DRAM chip performance due to FPGA's lower operating frequency compared to ASICs. For the switch-attached CXL-ASIC, the measured latency is 3.75 times larger than that of DDR-L; the additional \SI{203}{\nano\second} delay compared to direct-attached CXL-ASIC is introduced by the CXL switch.

Notably, the latencies between \SI{128}{MB}  and \SI{384}{MB} for CXL-DMSim are consistently higher than those measured on real hardware. This discrepancy is primarily due to gem5’s LLC model not capturing the full complexity of real LLC behavior. Specifically, when the pointer-chase working set goes up above the LLC size (\SI{96}{MB}) but remains below four times (\SI{384}{MB}), the observed latencies reflect a mixed effect of LLC hits and misses. Because gem5’s LLC model per se exhibits a high error rate (recent work \cite{gem5Tune} reports up to 56.4\% modeling error versus just 6.2\% for its DRAM model), CXL-DMSim poorly reproduces the intermediate-range latency measured in real hardware. A secondary influential factor is that the simulated LLC size is \SI{1.5}{MB} smaller than that on our testbed due to gem5's LLC configuration constraints; a higher setting (e.g., \SI{98}{MB}) leads to a smaller gap.

When using remote NUMA to emulate CXL memory, the read latency is relatively lower compared to real CXL memory devices, being about 1.53 times higher than that of DDR-L. Clearly, the experimental results show that CXL-DMSim can accurately simulate the access latency of real CXL memory devices, while the remote NUMA emulation cannot reflect the real latency of the devices.

\begin{figure}[tb]
\includegraphics[width=0.5\textwidth]{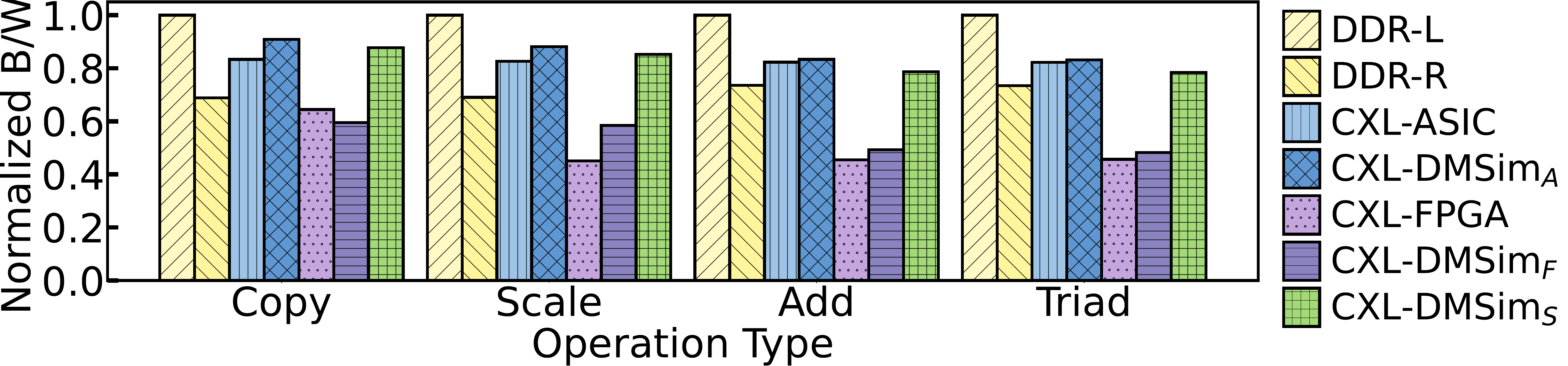}
\caption{Measured memory bandwidth normalized to DDR-L with copy, scale, add, and triad tests for different memory types.} 
\label{fig:STREAM_bw}
\end{figure}

\begin{figure*}[t]
\centering
\includegraphics[width=0.95\textwidth]{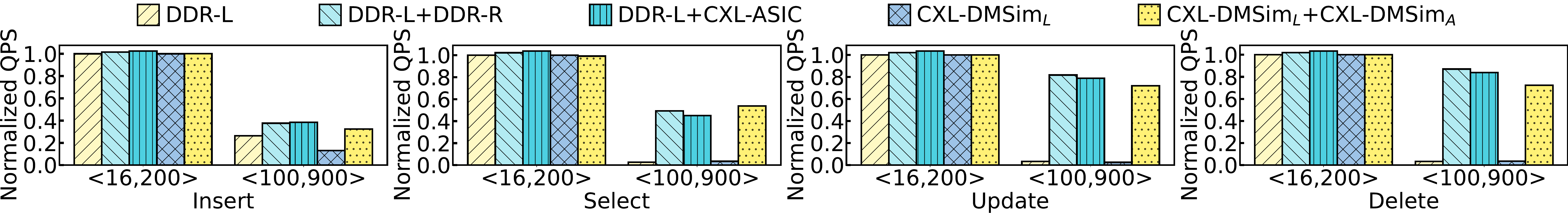}
\vspace{-5pt}
\caption{QPS of Viper with various memory-type expansion strategies for different database operations.}  
\label{fig:viper_qps}
\vspace{-5pt}
\end{figure*}

\subsubsection{\textbf{STREAM Test for Memory Bandwidth}}
To measure the sustainable memory bandwidth of our prototyped and simulated systems, we used STREAM \cite{stream} as a benchmark, which provides four tests, i.e. Copy, Scale, Add and Triad with different memory access and computing patterns. To ensure that STREAM measures realistic bandwidth between memory and processor without interference from caches, the data size has to be at least four times the size of LLC; thus we set the data size to \SI{1024}{MB} in our experiments. 

Fig.~\ref{fig:STREAM_bw} shows the measurement results of the seven different memory devices. Note that all the values are normalized to the bandwidth of DDR-L for the purpose of better comparison. It can be seen that the bandwidths derived from CXL-DMSim closely align with those obtained from the CXL hardware prototype; the modeling error rate is about 4.3\% on average. The CXL-FPGA memory achieves approximately 45\%-69\% of the DDR-L bandwidth. Comparatively, the CXL-ASIC memory exhibits better performance in bandwidth, achieving about 82\%-83\% of the DDR-L bandwidth. One can also see that the bandwidth of DDR-R is 68\%-74\% of the DDR-L bandwidth. This indicates that the bandwidth of DDR-R memory lies between CXL-ASIC and CXL-FPGA memories. The simulated switch-attached CXL-ASIC (CXL-DMSim\textsubscript{S}) shows similar bandwidth to the direct-attached CXL-ASIC.

In a nutshell, the relationship of the measured bandwidth is CXL-L$>$CXL-ASIC$>$DDR-R albeit the DDR5 bandwidth is the same in all three cases. This can be explained as follows. In the DDR-R case, UPI uses a directory-based home snoop coherency protocol \cite{4thintel_xeon}. A cache-line request from the local core to remote memory (DDR‑R) is first checked by the local \textit{caching and home agent} (CHA) to determine whether any copies exist within the NUMA node (by consulting the \textit{snoop filter} and LLC). In case none is found, it is forwarded to the remote (home) node’s CHA for cross‑node cache‑coherence checking.
The memory directory tracking remote cache-line state on DRAM chips and its on-die cache (HitME) \cite{hitme} are also involved to maintain coherence across NUMA nodes~\cite{moesi-prime}. Under STREAM tests with mixed read/write operations, the frequent cross-UPI coherence checks, especially on writes, become a bottleneck limiting DDR-R bandwidth \cite{exploreASIC}. In contrast, the CXL-ASIC memory is physically located at the same socket as the DDR-L memory, which does not involve cross-UPI cache coherency. The reason why the bandwidth of CXL-ASIC is smaller than DDR-L is owing to the overhead of packing and unpacking CXL protocol packets as well as the current immature design of the CXL memory controller~\cite{DasSharma2023AnIT}. Finally, CXL-FPGA exhibits the lowest bandwidth due to the memory-bandwidth limitations of its on-device DDR4 memory and the inherently lower operating frequency of the FPGA.

\begin{figure}[t]
\centering
\includegraphics[width=0.43\textwidth]{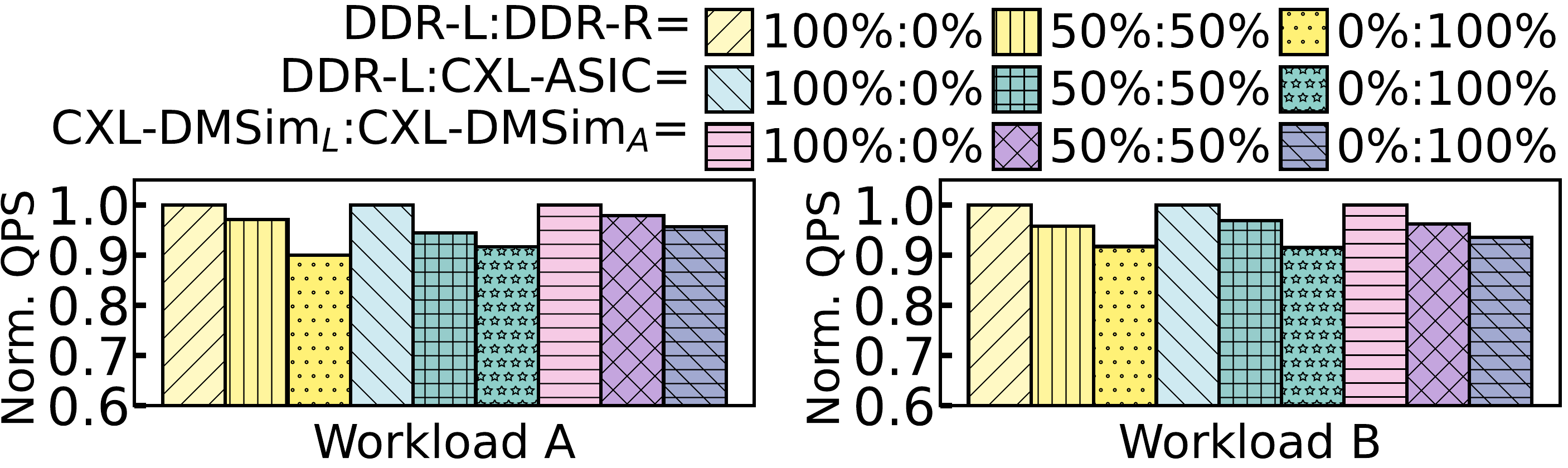}
\vspace{-5pt}
\caption{QPS of Redis with different ratios of pages allocated to DDR-L and DDR-R/CXL-HDM.} 
\label{Fig:Redis}
\end{figure}

\subsubsection{\textbf{Real-World Application Test}}
With the above characterization results, we can see that the performance of CXL-FPGA is much poorer than the real CXL-ASIC device. Due to the simulation overhead as well as page limitations, we will limit our real-world application experiments to CXL-ASIC only. To better understand the system performance of CXL-based disaggregated memory, we selected a representative in-memory database named Redis \cite{Redis} (version 5.0.5) to evaluate the system throughput in diverse application scenarios. The performance of Redis is evaluated using YCSB, a popular database benchmark suite \cite{ycsb_cooper2010benchmarking}. We used predefined workloads A and B, both following a Zipfian distribution. We employed YCSB’s default \SI{1}{KB} key-value size, an \SI{800}{MB} working set, and a single thread for the experiments.

With \emph{numactl}, we can direct Redis to allocate memory in DDR-L, DDR-R, CXL, or interleave between them to simulate a heterogeneous memory allocation strategy. Fig.~\ref{Fig:Redis} shows the \textit{queries per second} (QPS) of Redis under two workloads with different memory allocations. Using only DDR-L yields the highest QPS (42K QPS for Workload A and 39K QPS for Workload B), while interleaved allocations reduce QPS, and 100\% CXL or DDR-R leads to the lowest QPS. Additionally, an interleaving of DDR-L and DDR-R achieves slightly higher QPS than the same combination of DDR-L and CXL memories.
Redis is a memory-latency-sensitive application operating at \si{\mu\second}-level precision \cite{sun2023demystifying,exploreASIC}. For such applications, even a minor allocation of their working set to high-latency memory can lead to significant performance degradation. DDR-R, due to its moderate latency characteristics, may exhibit performance superior to that of CXL-ASIC.

\begin{mdframed}[style=MyFrame]
    {\textbf{Key takeaways:} (1) The performance characteristics of CXL memory expanders are remarkably different from that of DDR memories and they also vary significantly depending on  vendors and connection modes. 
    (2) CXL-DMSim can accurately and flexibly model real CXL memory devices with varying performance characteristics, while the NUMA-based emulator cannot.
    (3) The higher latency of CXL memories can degrade the performance of latency-sensitive applications (e.g., in-memory databases) and therefore should be deployed with caution.}
\end{mdframed}

\subsection{Exploration of CXL Memory Benefits}
\subsubsection{\textbf{Memory Capacity Expansion}}
Viper is a hybrid PMEM-DRAM key-value database \cite{viper}. We used Viper to evaluate the impact of memory expansion strategies on application performance when the local memory capacity is inadequate. To create this scenario meanwhile speeding up our experiments, we limited the DDR-L capacity accessible to Viper to ensure that it is less than the amount of data inserted into Viper. 
To evaluate the QPS of the system, we inserted an equal number of key-value pairs with two different types: \textless16,200\textgreater (0.216KB) and  \textless100,900\textgreater (1KB). The system employs a preferred memory allocation strategy, which attempts to allocate memory from DDR-L first. 

Fig.~\ref{fig:viper_qps} shows the test results of system performance in normalized QPS. When inserting key-value pairs of \textless16,200\textgreater, the QPS remains relatively consistent across various memory configurations. This is because the total data volume inserted into  Viper did not exceed the maximum capacity of DDR-L during the execution of the four operations. 
However, when the size of the inserted key-value pairs increased to \textless100,900\textgreater, the system's QPS with DDR-L alone dropped significantly. The decline is attributed to the total volume of inserted data  surpassing the capacity threshold that DDR-L can accommodate. It cannot completely accommodate the data intended to be inserted into Viper, necessitating frequent swapping of pages in and out using swap space by default. It can be seen that expanding memory with  CXL memory can enhance system throughput significantly with at most 25 times or 23 times improvement compared to DDR-L alone. 

\subsubsection{\textbf{Memory Bandwidth Expansion}}
To evaluate the benefits of CXL memory to bandwidth-sensitive applications, we performed inference tasks using Meta’s \emph{Deep Learning Recommendation Model} (DLRM) configured similar to the prior work MERCI \cite{lee2021merci}. We conducted tests with various memory mixing configurations, covering DDR-L, DDR-R, and CXL HDM with an interleaved memory allocation strategy.

\begin{figure}[tb]
\centering
\includegraphics[width=0.5\textwidth]{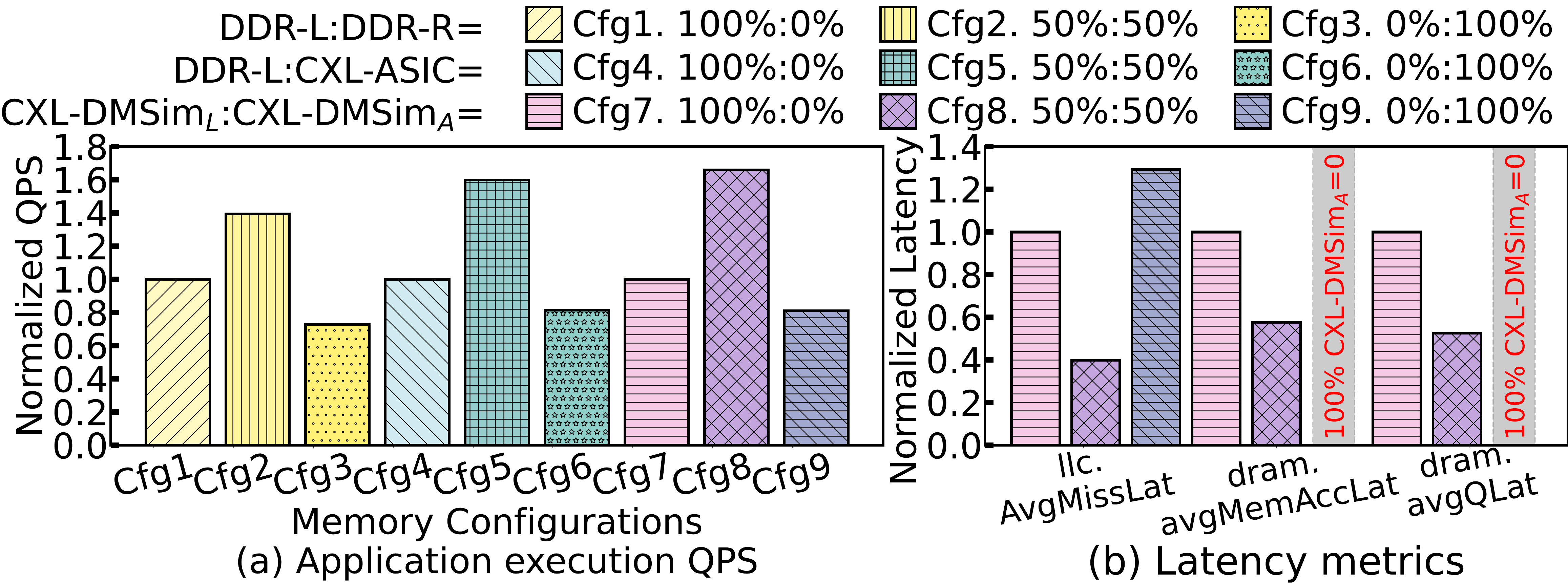} 
\vskip -5pt
\caption{Impact of memory mixing strategies on the inference performance of the DLRM model.}  
\label{fig:merci}
\end{figure}

Fig.~\ref{fig:merci}(a) presents the test results of DLRM inference in various memory mixing scenarios. One can see that the system performance improves by 39\% and 60\% respectively, when using a 50\%:50\% mix of DDR-L and DDR-R/CXL-HDM compared to the 100\% DDR-L case. This is because interleaved memory can effectively utilize the bandwidth of multiple memory devices, thereby increasing the overall system QPS. When using a 50\%:50\% mix of DDR-L and CXL HDM, the QPS of the system is higher than the QPS achieved with a 50\%:50\% mix of DDR-L and DDR-R. This improvement stems from the superior bandwidth of CXL-ASIC memory over DDR-R. Notably, CXL-DMSim produced simulation results consistent with real hardware observations, demonstrating its capability to accurately model the characteristics of CXL memory devices. To gain deeper insights into the root causes of bandwidth expansion in interleaved memory, we analyzed detailed statistics collected from the simulator. In Fig.~\ref{fig:merci}(b), the metric \emph{llc.AvgMissLat} measures the average LLC miss latency, \emph{dram.avgMemAccLat} represents the average memory access latency per DRAM burst, and \emph{dram.avgQLat} quantifies the average queuing latency per DRAM burst in DDR memory. Note that in the CXL-ASIC exclusive configuration, there is no DDR traffic; thus, both \emph{dram.avgMemAccLat} and \emph{dram.avgQLat} are zero. The statistical results demonstrate that the interleaved DDR and CXL memory configuration consistently exhibited the lowest latency across all metrics. This improvement is primarily due to CXL memory offloading half of the memory traffic, significantly reducing queuing delays and access contention. Consequently, this configuration achieved the lowest LLC average miss latency.

\begin{mdframed}[style=MyFrame]
    {\textbf{Key takeaways:} CXL memory arises with benefits of on-demand expanding system memory capacity and bandwidth, which greatly boost the performance of applications that are memory capacity-bound and/or bandwidth-bound.}
\end{mdframed}

\subsection{Congestion Analysis Using CXL-DMSim}
During our experiments, we observed an interesting phenomenon in both  CXL-DMSim and the hardware platform. After a certain point, the QPS for MERCI drops as the number of cores continuously increases. Using the performance counter monitor (PCM) \cite{pcm} on the hardware testbed, we found that the average LLC miss latency in the 48-core system was \SI{2.3}{\micro\second}, compared to \SI{568}{\nano\second} in the 12-core system. Due to the limited observability of the real hardware, we used CXL-DMSim to dig into the root causes. We collected key statistics for both the simulated 12-core and 48-core systems, as listed in Table~\ref{tab:statistics}. Stats.1 shows that the aggregate QPS of the 48-core system is only 69\% of that of the 12-core system. Meanwhile, Stats.2 and Stats.3 reveal that both the average per-core load-to-use latency and the average LLC miss latency are higher in the 48-core system than in the 12-core system, highlighting the increased latency along the CXL memory access path in the 48-core configuration. Stats.4 records the count of blocked events of LLC requests caused by miss status handling register (MSHR) entry exhaustion. As MSHRs are critical for managing concurrent cache miss requests, their depletion blocks subsequent CPUs from continuing to issue memory access requests, significantly degrading system performance.

\begin{table}[t]
\centering
\renewcommand{\arraystretch}{1.0} 
\caption{Statistics from 12-core and 48-core DLRM experiments\\on the CXL-DMSim\textsubscript{A} memory (Cfg9).}
\label{tab:statistics}
\resizebox{0.49\textwidth}{!}{
\begin{tabular}{|c|l|l|l|}
\hline
\textbf{No.} & \textbf{Statistics}  & \textbf{12-Core}  & \textbf{48-Core}   \\ 
\hline
{1} & {Aggregate QPS} & {1.6e6}  & {1.1e6} \\ \hline
{2} & {core.loadToUse::mean (Cycle)}& {97}  & {143} \\ \hline
{3} & {llc.overallAvgMissLat. (Tick)} &{3.1e5} &{9.6e5}\\ \hline
{4} & {llc.mshrBlocked (Count)} &{0} &{1.6e7}\\ \hline
{5} & {bridge.avgReqQueOccupancy} &{22.6\%} &{43.0\%}\\ \hline
{6} & {bridge.avgRspQueOccupancy} &{48.7\%} &{99.8\%}\\ \hline
{7} & {cxlCtrl.avgReqQueOccupancy} &{13.6\%} &{13.9\%}\\ \hline
{8} & {cxlCtrl.avgRspQueOccupancy} &{66.3\%} &{99.0\%}\\ \hline
{9} & {cxlMem.avgRdBufOccupancy} &{14.4\%} &{21.9\%}\\ \hline
{10} & {cxlMem.avgWrBufOccupancy} &{37.8\%} &{38.2\%}\\ \hline
\end{tabular}
}
\end{table}

To further analyze the root cause, we conducted an in-depth examination of packet behaviors along the CXL transmission path, which consists of four components: the bridge, the CXL controller, memory controller, and the memory medium. In our CXL link implementation, reliable sequential transmission and credit-based flow control \cite{cho2017credit} follow two key constraints: 1) a requester must ensure the receiver's request queue has sufficient capacity (credits) before sending packets; 2) for requests requiring responses, the receiver must pre-allocate enough space in the response queue to prevent deadlock (e.g., deadlock occurs when subsequent requests depend on the results of preceding requests, but these responses cannot be returned due to a full response queue, thereby blocking further requests). Stats.5-10 quantify the average resource utilization of different components along the CXL transmission path. Compared to the 12-core system, we observed severe packet queuing congestion in the response queues of both the bridge and the CXL controller in the 48-core system. Meanwhile, the occupancy of the read/write buffers in the backend memory module, though slightly higher, remains below 40\% of the full capacity. These observations suggest that the congestion primarily originates from the CXL controller rather than the  memory module. Specifically, in the 48-core system, highly concurrent memory accesses rapidly depleted the available credits for the response queue in the CXL controller. This congestion then causes back-pressure on the bridge and LLC, ultimately exhausting LLC MSHR entries and thus blocking further CPU memory accesses, thereby degrading system QPS.

\begin{mdframed}[style=MyFrame]
    {\textbf{Key takeaways:} Through hardware measurements and simulations, we identified a congestion phenomenon and its root cause in the CXL memory device under high-concurrency workloads, revealing that the CXL controller’s concurrent processing capability is a key factor affecting CXL memory throughput. This exploration underscores the high configurability and observability advantages of CXL-DMSim, enabling researchers to precisely pinpoint system bottlenecks and flexibly explore optimization strategies.}
\end{mdframed}

\vspace{-10pt}
\subsection{Expandability of CXL-DMSim}

\begin{figure}[tb]
\centering
\includegraphics[width=0.32\textwidth]{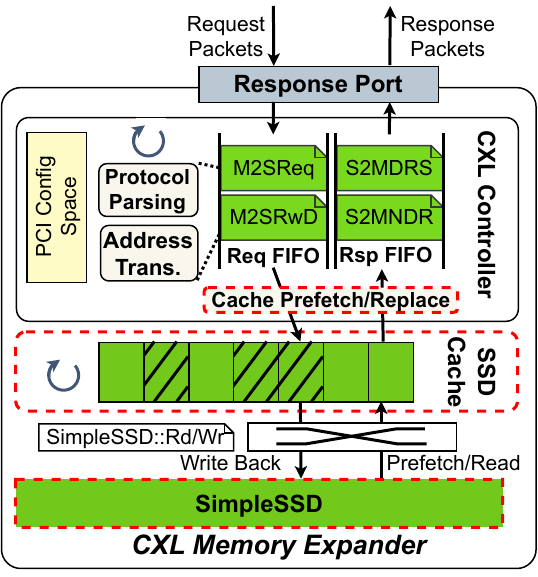}
\caption{CXL-SSD device model design.} 
\label{fig:CXL_SSD_model}
\end{figure}

CXL-DMSim adheres to the memory technology-agnostic principle of the CXL protocol, enabling seamless support for a rang of media such as DRAM, Flash, and emerging NVMs. As an example, Fig. \ref{fig:CXL_SSD_model} shows the device model of CXL-SSD, which is a newly added component to CXL-DMSim.
The red dashed blocks highlight the added modules to our CXL device model. We utilized an open-source SSD simulator called SimpleSSD \cite{simpleSSD_paper} to model the SSD backend of the CXL interface. However, two issues must be addressed. First, SSD exhibits significantly higher latency (\si{\micro\second} level) compared to DRAM (\si{\nano\second} level). Second, CXL.mem is based on memory semantics at a byte granularity, while SSD operates with I/O semantics at a page (e.g., 4KB) granularity. 
A simple replacement of the memory medium from DRAM to SSD would inevitably result in a dramatic performance decline. Thus, we designed a cache between the CXL controller and SSD model to address the aforementioned issues.
Additionally, we implemented a cache prefetching and replacement module to manage cache lines. Specifically, we first considered that, due to the larger access granularity of SSDs, prefetching is more likely to result in cache pollution. Our prefetching decisions are based on the history of cache misses, with prefetching only executed when there is a high confidence level that it will yield benefits. Furthermore, given the higher access latency of SSDs, we employ the Best-Offset prefetching algorithm \cite{michaud2016BO}, which prioritizes timeliness in retrieving data from the backend SSD. When a memory access request reaches the CXL-SSD, it is directly returned if the SSD cache hits; otherwise, it is translated into SimpleSSD data packets for backend retrieval.

\begin{figure}[tb]
\centering
\includegraphics[width=0.38\textwidth]{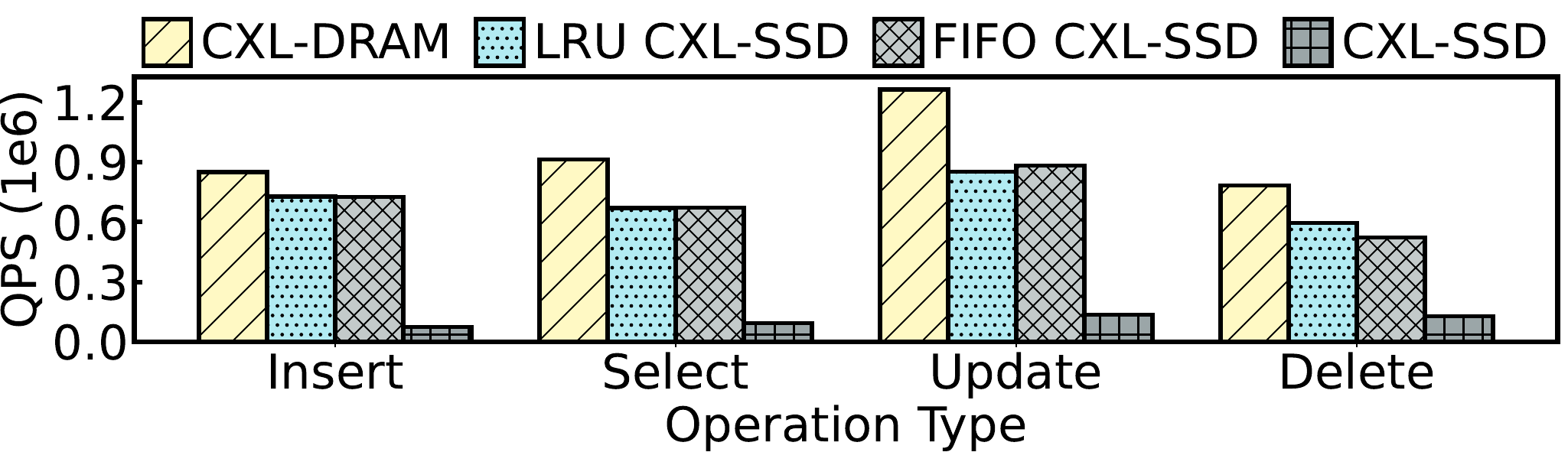}
\vspace{-5pt}
\caption{QPS of Viper with CXL-DRAM and CXL-SSD for four operations.} 
\label{fig:Viper_CXLSSD}
\end{figure}

We ran the Viper test on the CXL-SSD to evaluate its performance. Unlike the previous tests, we utilized the AM mode in this case to demonstrate its usage. As illustrated in Fig.~\ref{fig:Viper_CXLSSD}, the test results indicate that the QPS of CXL-SSD across four operations are considerably lower than CXL-DRAM, primarily due to the higher access latency inherent in the Flash medium. To evaluate the effects of cache on system performance, we implemented and compared different cache replacement policies including least recently used (LRU) and first in first out (FIFO). It can be seen that the QPS of LRU or FIFO CXL-SSD with an additional cache has been significantly enhanced, reaching 72\%-88\% of CXL-DRAM.  
These results not only prove the effectiveness of the AM mode, but also demonstrate the high expandability of our simulator.

Nevertheless, the design of CXL-SSD as a part of memory pool still faces many challenges and trade-offs that warrant further investigation: \textbf{(1) Bridging the performance gap between SSD and DRAM.} The inherent high latency of flash media can severely degrade system performance, thus innovations are required to boost CXL-SSD performance (e.g., device-side cache, dedicated prefetch strategies). Moreover, designing an efficient tiered memory system is also pivotal for the practical deployment  of CXL-SSD. \textbf{(2) Coordination of CXL.io and CXL.mem transfers.} Although CXL.mem provides lower-latency, memory-semantic access to the SSD, the DMA-based block transfer means supported by CXL.io must not be abandoned since it is superior in transferring data blocks larger than \SI{1}{KB} \cite{demystifying_type2}. Efficiently coordinating both transfer channels at the software level remains a significant challenge. \textbf{(3) Volatile versus non-volatile usage modes.} 
CXL-SSD should not be simply used as a type of cheap and large-volume  memory complementing the traditional volatile main memory, its persistent property can make a big difference. 
CXL-SSD and more broadly CXL-NVM bring a revival of persistent memory, which partly replaces the storage layer thus is vital to applications such as in-memory databases, big data analytics, and virtualization.

\begin{table*}[tb]
\centering
\renewcommand{\arraystretch}{1} 
\caption{Comparison between CXL-DMSim and other CXL system simulators: pros \& cons.}
\label{tab:cxldmsim_vs_others}
\resizebox{1\textwidth}{!}{
\begin{tabular}{|l|c|c|c|c|c|c|c|c|}
\hline
\textbf{Attribute}  & \textbf{CXL-DMSim }  & \textbf{QEMU} & \textbf{Mess+gem5} & \textbf{gem5-cxl} & \textbf{DisaggSim} & \textbf{DRackSim} & \textbf{CXLMemSim} & \textbf{Remote NUMA}  \\ \hline
{Reference} & {This work}& {\cite{QEMU}}&{\cite{mess}}&{\cite{gem5-CXL-github}} &{\cite{DisaggSim}} & {\cite{DRackSim}}&{\cite{CXLMemSim_github}}&{\cite{NUMA_sim1,arif2022exploiting}}\\ \hline
{CXL protocol support} & {1.1+}& {2.0}&{No}&{No} &{No} & {No} & {No} & {No}\\ \hline
{Full-system CXL support} & {Yes}& {Yes}&{No}&{No} &{No} & {No} & {No} & {No}\\ \hline
{Silicon validation} & {Yes}& {No}&{No}&{No} &{No} & {No} & {No} & {N/A}\\ \hline
{Expandability} & {High}& {High}&{Low}&{Low} &{Medium} & {Medium} & {Low} & {Low}\\ \hline
{Sw/hw co-design} & {Yes}&{Yes}&{No}&{No}&{No}&{No} & {No} & {No}\\ \hline
{Cycle accuracy} & {Yes}& {No}&{Yes}&{Yes} &{Yes} & {Yes} & {No} & {Yes}\\ \hline
{Development maturity} & {Yes}& {Yes}&{Yes}&{No} &{Yes} & {Yes} & {No} & {Yes}\\ \hline
{Configurability} & {High}& {High}&{High}&{Low} &{High} & {High} & {Medium} & {Low}\\ \hline
{Simulation error} & {Low}& {High}&{Medium}&{High} &{High} & {High} & {Unknown} & {High}\\ \hline
{Simulation speed} & {Low}& {Medium}&{Low}&{Low} &{Low} & {Low} & {Medium} & {High}\\ \hline
\end{tabular}
}
\vspace{-10pt}
\end{table*}

\section{Discussion}
\label{sec:Discussion}

\begin{figure}[tb]
\includegraphics[width=0.45\textwidth]{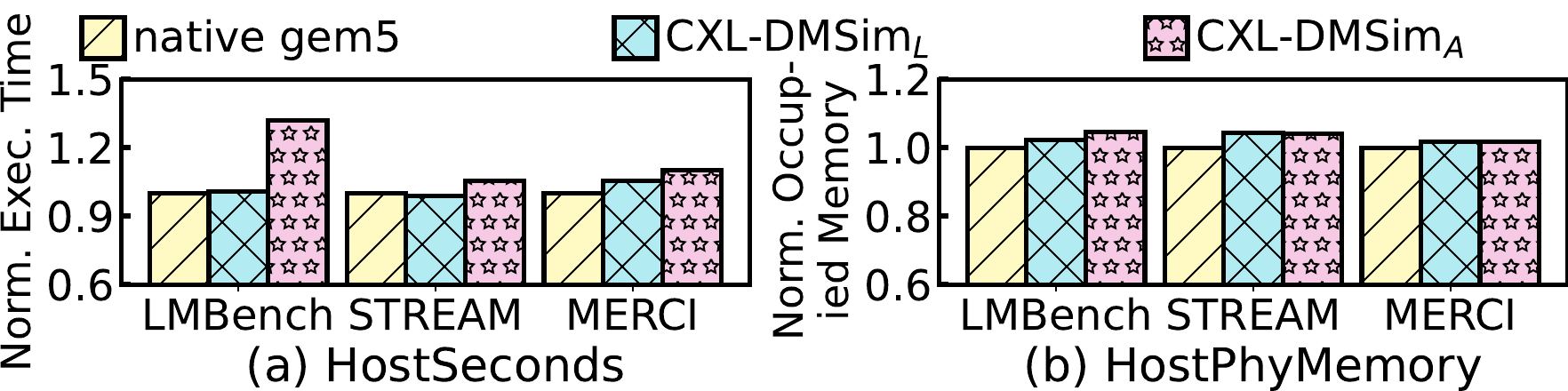}
\caption{Simulation overhead in terms of simulation speed and occupied host memory for CXL-DMSim.} 
\label{fig:CXL-DMSim_overhead}
\end{figure}

\subsection{Simulation Overhead of CXL-DMSim}
To assess the simulation overhead covering two aspects which are \textit{simulation speed} and \textit{occupied host memory} introduced by CXL-DMSim in comparison to the native gem5, we conducted various experiments on our simulator and collected statistics of the execution time (HostSeconds) and physical memory occupation (HostPhyMemory) on the host machine. Fig. \ref{fig:CXL-DMSim_overhead}(a) compares the HostSeconds between native gem5, CXL-DMSim\textsubscript{L}, and CXL-DMSim\textsubscript{A} with respect to three workloads.
The execution time of CXL-DMSim\textsubscript{L} is nearly identical to that of native gem5, while the average execution time of CXL-DMSim\textsubscript{A} is 16\% higher. This is because CXL-DMSim\textsubscript{A} involves more components and longer latency in the simulation process. Notably, LMbench, a benchmark focused on evaluating memory access latency, exhibits slightly higher execution times compared to STREAM and MERCI.

Fig.~\ref{fig:CXL-DMSim_overhead}(b) compares HostPhyMemory overhead. 
CXL-DMSim exhibits an average increase of only 3.5\% in host memory usage, compared to raw gem5. This increase results from the additional memory allocation necessary on the host machine to model the CXL device memory. Moreover, the magnitude of memory overhead is closely correlated with the amount of physical memory utilized by different applications.

\begin{figure}[tb]
\centering
\includegraphics[width=0.45\textwidth]{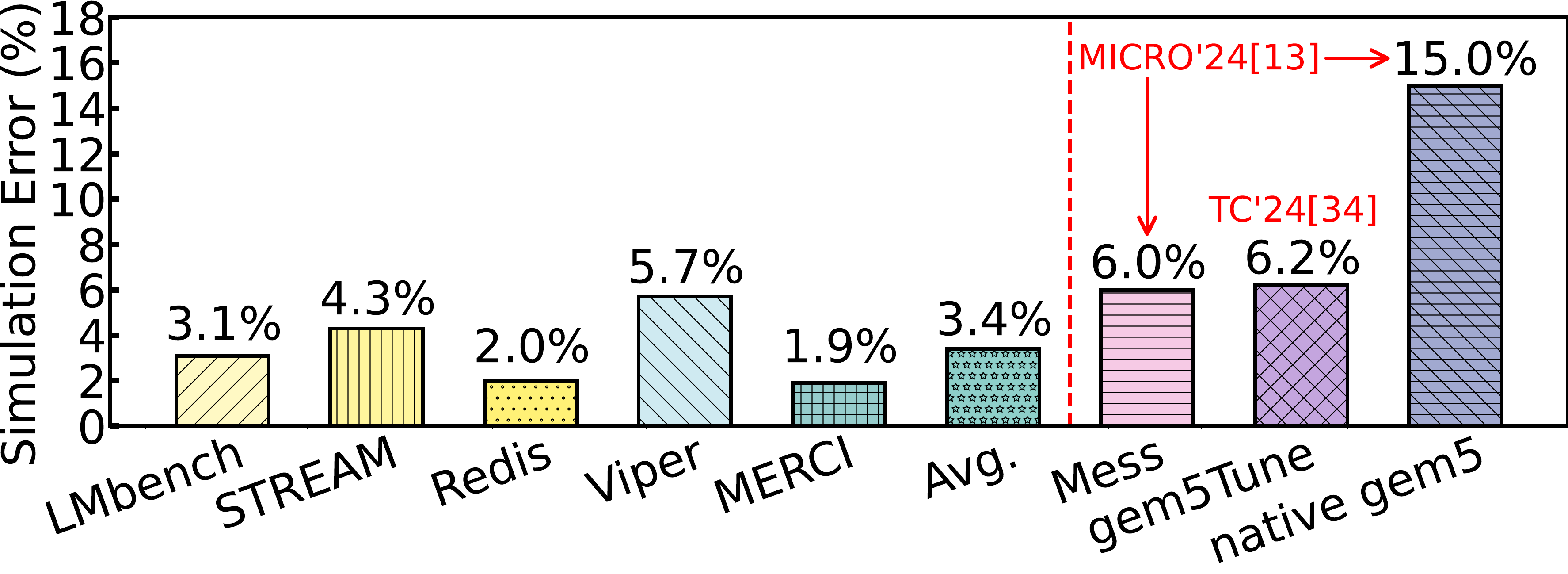}
\vspace{-5pt}
\caption{Simulation error of our and related simulators in the prior art.} 
\label{fig:CXL-DMSim_simulation_error}
\end{figure}

\vspace{-5pt}
\subsection{Simulation Error of CXL-DMSim}
We also analyzed the simulation error stats of CXL-DMSim for a large variety of workloads against the results on our hardware testbed. The collected error data is compared with that of other CXL simulators as depicted in Fig.~\ref{fig:CXL-DMSim_simulation_error}. It can be seen that the minimum simulation error is 1.9\% for MERCI, while the maximum goes to Viper. The averaged simulation error is measured at 3.4\% across all workloads in our experiments. CXL-DMSim exhibits the best accuracy in comparison to the two latest works Mess (6.0\%) and gem5Tune (6.2\%) as revealed in their respective papers \cite{mess, gem5Tune}. The native gem5 shows the highest average error of 15.0\% \cite{mess}. 
 
The sources of simulation errors for CXL-DMSim can be attributed to the cumulative effects of modeling inaccuracies in the entire system. These include but are not limited to the following. (1) The CPU model lags behind the latest CPU we used on our testbed. (2) The cache model accounts for a large part of errors,  particularly when data volumes range from one to four times the LLC capacity. (3) The poor modeling of the disk in gem5 when storage layers are involved, as can be observed with the increased errors for Viper when some data was pushed out to disk. Note that it is always a dilemma in the pursuit of simulation accuracy and speed.

\vspace{-5pt}
\subsection{CXL-DMSim vs. Other Simulators}
Table~\ref{tab:cxldmsim_vs_others} compares CXL-DMSim with other prominent simulators at various aspects. 
First, CXL-DMSim and QEMU offer comprehensive support for the CXL protocol, which is critical for accurately simulating the behavior of real CXL memory systems. In contrast, the other six simulators actually do not support the CXL protocol. Notably, although DisaggSim and DRackSim support multi-node interconnection simulation, they lack of modeling the CXL protocol's specific workflow and device behavior. Second, CXL-DMSim is a full-system CXL simulator that provides the most realistic interaction between the operating system and simulated hardware, opening up a range of possibilities for software-hardware co-design. In contrast, Mess, gem5-CXL, CXLMemSim, DisaggSim, and DRackSim cannot boot the operating system that supports CXL devices.
Third, CXL-DMSim is the only one which has undergone extensive silicon validation, surpassing the other simulators in terms of accuracy and realism. CXL-DMSim also excels in expandability, cycle accuracy, development maturity, and configurability with acceptable simulation error. Nevertheless, these advantages come along at the cost of slow simulation speed, high modeling complexities for new features, and limited system scale. 

\vspace{-5pt}
\subsection{Future Work}
CXL-DMSim now fully supports the CXL.io, CXL.cache, and CXL.mem sub-protocols, and multiple use cases for type-3 and type-1 devices have been implemented. Owing to space limitations, the calibration and implementation details of CXL.cache, along with related use cases (CXL-NIC), will be presented in another paper. Incremental support for type-2 use cases (CXL-GPU) is currently underway. Moreover, a single-host CXL switch has been realized to interconnect multiple CXL devices. Leveraging this foundation, we are progressing towards a distributed CXL-DMSim simulation platform, analogous to dist-gem5 \cite{distgem5}. In this framework, compute nodes, memory nodes, and switch-based fabric interconnect execute within distinct processes; packet exchanges and clock synchronization among these heterogeneous simulation nodes are facilitated through shared memory. Ultimately, this architecture enables a large-scale disaggregated memory system built upon a configurable CXL fabric.

\section{Conclusion}
\label{Sec:Conclusion}
In this paper, we have presented CXL-DMSim, an open-source and silicon-calibrated full-system simulator for CXL disaggregated memory systems. CXL-DMSim can be used either in a NUMA-compatible kernel-managed mode or in an app-managed mode depending on users' preference. Our experiments on both hardware and CXL-DMSim suggest that CXL memory is very effective in memory expansion of both capacity and bandwidth to boost system performance. With CXL-DMSim, the system performance bottlenecks can be quickly identified thanks to its high observability with substantial simulation statistics. In the future, we will continue to enhance the capabilities of CXL-DMSim, and we also welcome the community to join us in building a solid simulation platform for architectural research on memory-pooled computing systems.

\bibliographystyle{IEEEtran}
\bibliography{sample-base}

\newpage
\begin{IEEEbiography}[{\includegraphics[width=1in,height=1.25in,clip,keepaspectratio]{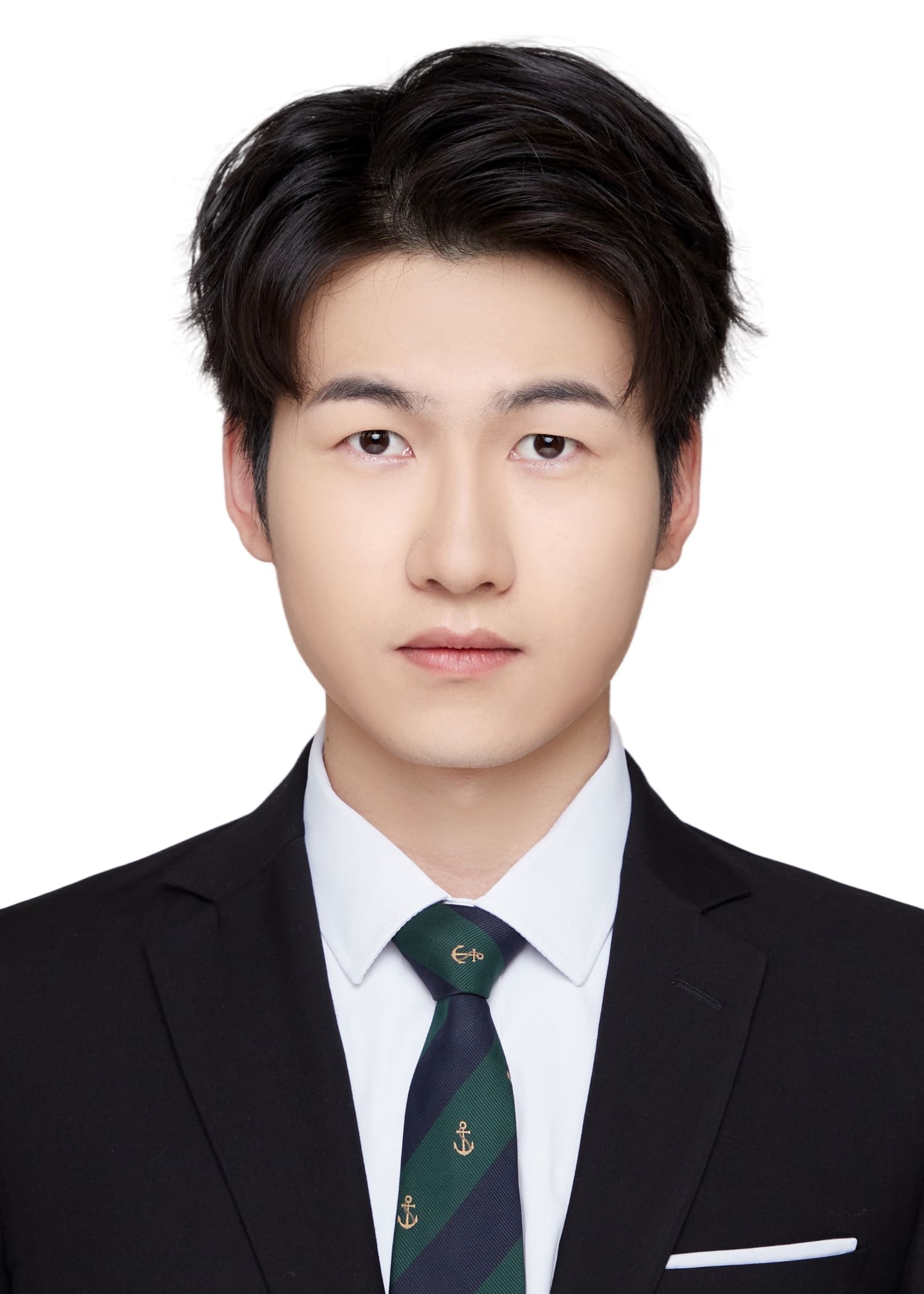}}]{Yanjing Wang} received the B.E. degree in computer science and technology from the College of Computer Science, Chongqing University, Chongqing, China, in 2023. He is currently pursuing the Ph.D. degree in computer science and technology from the College of Computer Science and Technology, National University of Defense Technology, Changsha, China. His current research interests are coherent heterogeneous computing and memory pooling systems.
\end{IEEEbiography}
\vskip -2\baselineskip plus -1fil
\begin{IEEEbiography}
	[{\includegraphics[width=1in,height=1.25in,clip,keepaspectratio]{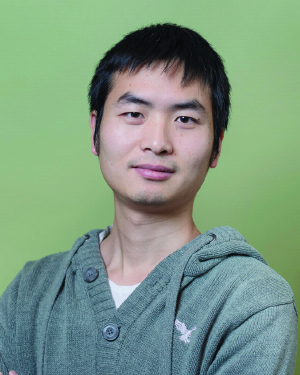}}]{Lizhou Wu} received the B.Sc. degree in electronic science and engineering from Nanjing University, the M.Eng. degree in computer science and technology from National University of Defense Technology (NUDT), and the Ph.D.  degree (Hons.) in computer engineering from Delft University of Technology. He is currently an assistant professor at the College of Computer Science and Technology, NUDT. His research interests cover three domains:  1) Coherent heterogeneous computing and memory pooling systems, 2) Spintronic design, automation, and test, 3) Emerging computing paradigms based on non-volatile memory devices. Dr. Wu has received many international awards such as IEEE TTTC's E.J. McCluskey Doctoral Thesis Award in 2021, Best Paper Award at DATE'20, and three  best paper nominations.
\end{IEEEbiography}
\vskip -2\baselineskip plus -1fil
\begin{IEEEbiography}
	[{\includegraphics[width=1in,height=1.25in,clip,keepaspectratio]{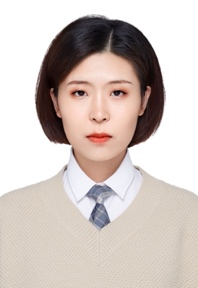}}]{Wentao Hong} received the B.E. degree in computer science and technology from the College of Computer Science and Technology, Jilin University, Changchun, China, in 2022, where she is currently pursuing the M.E. degree in computer technology from the College of Computer Science and Technology, National University of Defense Technology, Changsha, China. Her current research interests include memory pooling and congestion control.
\end{IEEEbiography}
\vskip -2\baselineskip plus -1fil
\begin{IEEEbiography}
	[{\includegraphics[width=1in,height=1.25in,clip,keepaspectratio]{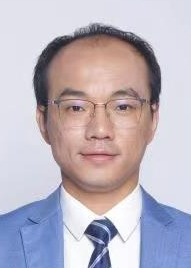}}]{Yang Ou} received the Ph.D. degree in computer science and technology from the National University of Defense Technology (NUDT), Changsha, China, in 2013. He is currently an associate professor at the College of Computer Science and Technology, NUDT. His research interests include computer architecture and high-performance interconnection technology.
\end{IEEEbiography}
\vskip -2\baselineskip plus -1fil

\begin{IEEEbiography}
	[{\includegraphics[width=1in,height=1.25in,clip,keepaspectratio]{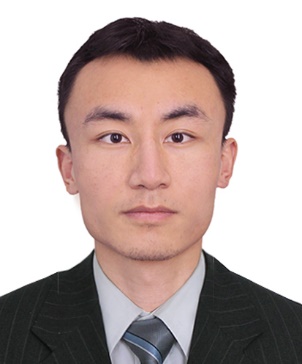}}]{Zicong Wang} received the B.S., M.S., and PhD degrees from the National University of Defense Technology (NUDT), Changsha, China, in 2008, 2014, and 2019, respectively. He is currently an assistant professor at the College of Computer Science and Technology, NUDT. His main research interests include networks-on-chip design and computer architecture.
\end{IEEEbiography}
\vskip -2\baselineskip plus -1fil
\begin{IEEEbiography}
	[{\includegraphics[width=1in,height=1.25in,clip,keepaspectratio]{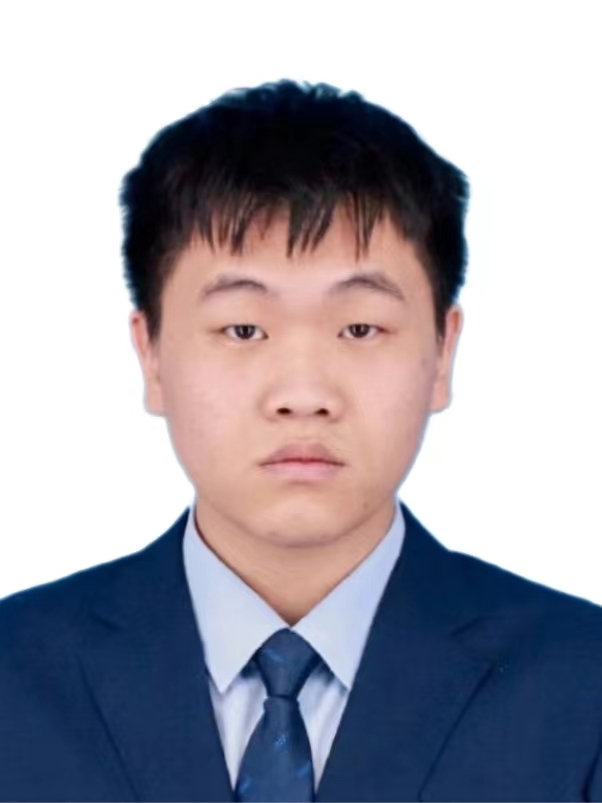}}]{Sunfeng Gao} received the B.E. degree in integrated circuit design and integrated systems from the College of Computer Science and Technology, National University of Defense Technology (NUDT), Changsha, China, in 2024. He is currently pursuing the Ph.D. degree in integrated circuit engineering from the College of Computer Science and Technology, NUDT. His current research interests are coherent heterogeneous computing and memory pooling systems.
\end{IEEEbiography}
\vskip -2\baselineskip plus -1fil
\begin{IEEEbiography}
	[{\includegraphics[width=1in,height=1.25in,clip,keepaspectratio]{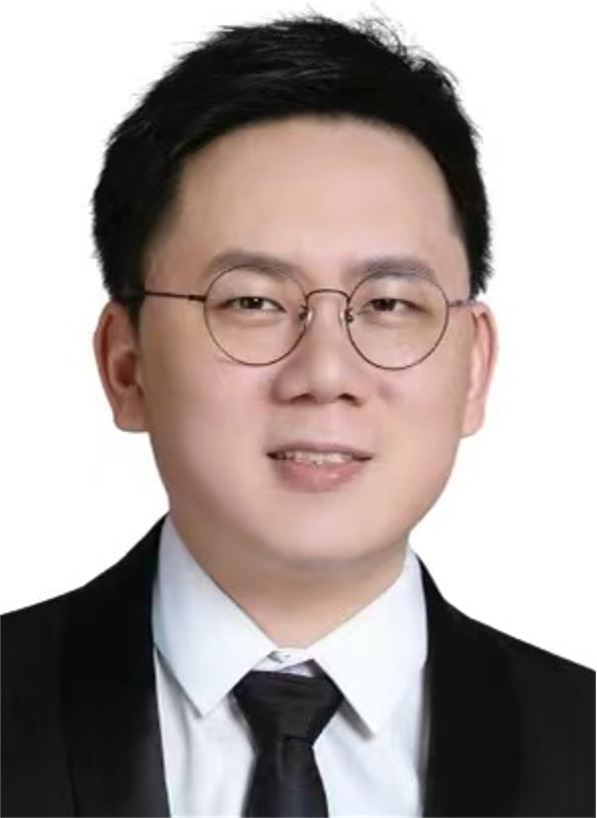}}]{Jie Zhang} received the Ph.D. degree from the Yonsei University, Incheon, Korea in 2020. He is currently an assistant professor of the school of Computer Science at Peking University, China. He is engaged in the research and design of storage systems, non-volatile memory and specialized processors. His research addresses the requirements for high-performance storage systems in the era of big data and artificial intelligence from the perspective of computer architecture.
\end{IEEEbiography}
\vskip -2\baselineskip plus -1fil
\begin{IEEEbiography}
	[{\includegraphics[width=1in,height=1.25in,clip,keepaspectratio]{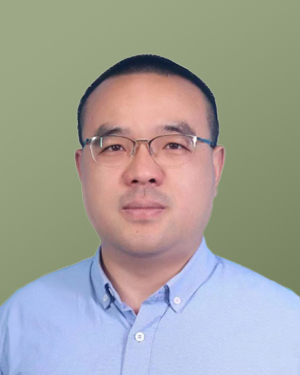}}]{Sheng Ma} received the B.S. and Ph.D. degrees in computer science and technology from the National University of Defense Technology (NUDT) in 2007 and 2012, respectively. He visited the University of Toronto from 2010 to 2012. He is currently a professor at the College of Computer Science and Technology, NUDT. He has authored over 30 papers in internationally recognized journals and conferences. His research interests include on-chip networks, SIMD architectures, and arithmetic unit designs.
\end{IEEEbiography}
\vskip -2\baselineskip plus -1fil
\begin{IEEEbiography}
	[{\includegraphics[width=1in,height=1.25in,clip,keepaspectratio]{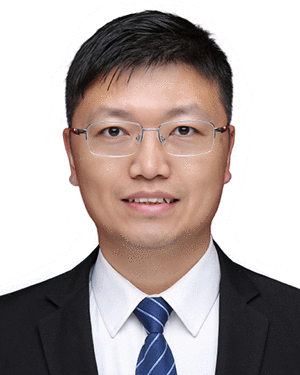}}]{Dezun Dong} received the B.S., M.S., and Ph.D. degrees from the National University of Defense Technology (NUDT), Changsha, in 2002, 2004, and 2010, respectively. Currently, he is a Professor with the College of Computer, NUDT, where he leads the research group of high-performance network and architecture (HiNA) and is also the Deputy Director-Designer of the Tianhe supercomputer. He has authored or co-authored more than 60 peer-reviewed papers in reputed international journals and conferences. His research interests include high-performance networks and architecture for supercomputers, data centers, and deep learning systems.
\end{IEEEbiography}
\vskip -2\baselineskip plus -1fil
\begin{IEEEbiography}
	[{\includegraphics[width=1in,height=1.25in,clip,keepaspectratio]{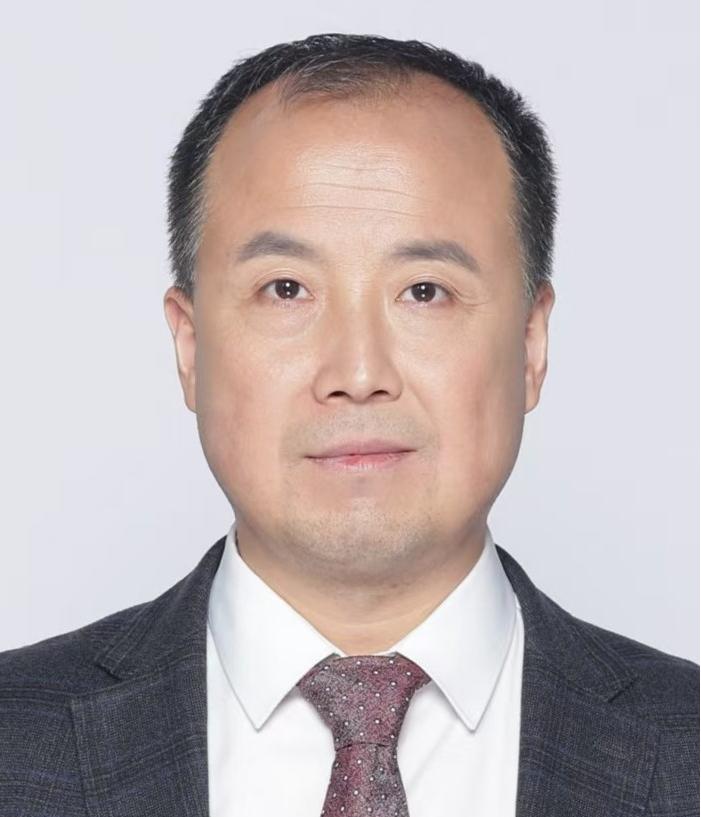}}]{Xingyun Qi} received the B.S., M.S., and Ph.D. degrees from the National University of Defense Technology (NUDT), in 2001, 2003, and 2009, respectively. He is currently an associate professor at the College of Computer Science and Technology, NUDT. His research interests include high-performance computer architecture, high-speed interconnection, and ASIC chip design.
\end{IEEEbiography}
\vskip -2\baselineskip plus -1fil
\begin{IEEEbiography}
	[{\includegraphics[width=1in,height=1.25in,clip,keepaspectratio]{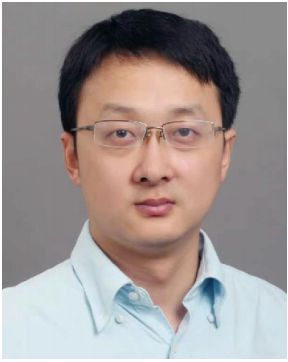}}]{Mingche Lai} received the B.S., M.S., and Ph.D. degrees in electrical engineering and computer science from the National University of Defense Technology (NUDT), in 2003, 2005, and 2008, respectively. He is currently a professor and the Chief Engineer at the College of Computer Science and Technology, NUDT. His research interests include high-performance interconnection, the hybrid integrated optoelectronic chip, HMC memory controller, and GPU architecture design. 
\end{IEEEbiography}
\vskip -2\baselineskip plus -1fil
\begin{IEEEbiography}
	[{\includegraphics[width=1in,height=1.25in,clip,keepaspectratio]{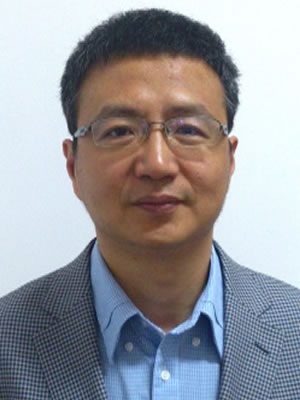}}]{Nong Xiao} (Senior Member, IEEE) received the Ph.D. degree in computer science and technology from the National University of Defense Technology (NUDT), Changsha, China, in 1996. He has been a professor since 2004 with the College of Computer Science and Technology, NUDT. He is a recipient of the National Science Fund for Distinguished Young Scholars and has been appointed the Changjiang Distinguished Professorship. He currently serves as the Chair of the Technical Committee on Information Storage Technology of the China Computer Federation (CCF). His research interests include computer architecture, high-performance computing, embedded systems, grid computing, and large-scale storage systems.
\end{IEEEbiography}

\end{document}